\documentclass[12pt]{article}

\pdfoutput=1

\usepackage{graphicx}
\usepackage{epsfig}
\usepackage{amsfonts}
\usepackage{amssymb}
\usepackage{amsmath}
\usepackage{dsfont}
\usepackage{bbm}
\usepackage{cite}
\usepackage{multirow}
\usepackage{colortbl}

\def\beq{\begin{equation}}
\def\eeq{\end{equation}}
\def\beqa{\begin{eqnarray}}
\def\eeqa{\end{eqnarray}}

\def\beq{\begin{equation}}
\def\eeq{\end{equation}}
\def\beqa{\begin{eqnarray}}
\def\eeqa{\end{eqnarray}}

\def\eps{{\epsilon }}

\def\cW{{\mathcal W}} 
\def\cK{{\mathcal K}}
\def\cG{{\mathcal G}}
\def\cP{{\mathcal P}}

\def\cZ{{\mathcal Z}}
\def\cz{{\mathcal Z}}
\def\cS{{\mathcal S}}
\def\cT{{\mathcal T}}

\def\wt{\widetilde}

\def\fg{{\mathfrak g}}

\def\a{\alpha}
\def\b{\beta}
\def\g{\gamma}
\def\d{\delta}

\def\VIIA{V_{\textrm{\begin{scriptsize}IIA\end{scriptsize}}}}

\def\Vloc{V_{\textrm{\begin{scriptsize}loc\end{scriptsize}}}}
\def\VGen{V_{\textrm{\begin{scriptsize}gen\end{scriptsize}}}}
\def\VT{V_{\textrm{\begin{scriptsize}T-fold\end{scriptsize}}}}

\def\VNS5{V_{\textrm{\begin{scriptsize}NS5\end{scriptsize}}}}
\def\3loc{V_{\textrm{\begin{scriptsize}O3/D3\end{scriptsize}}}}
\def\7loc{V_{\textrm{\begin{scriptsize}O7/D7\end{scriptsize}}}}
\def\6loc{V_{\textrm{\begin{scriptsize}O6/D6\end{scriptsize}}}}

\def\fnG{\textrm{\begin{footnotesize}G\end{footnotesize}}}
\def\fnNG{\textrm{\begin{footnotesize}NG\end{footnotesize}}}

\begin{document}
\pagestyle{plain}

\makeatletter
\@addtoreset{equation}{section}
\makeatother
\renewcommand{\theequation}{\thesection.\arabic{equation}}
\pagestyle{empty}
\rightline{IFT-UAM/CSIC-09-36}
\vspace{5mm}
\begin{center}
\LARGE{\bf
 Flux moduli stabilisation, Supergravity algebras and no-go theorems
\\[5mm]}
\large{
Beatriz de Carlos${}^{a}$,  Adolfo Guarino${}^{b}$ and Jes\'us M. Moreno${}^{b}$
\\[3mm]}
\small{
${}^a$
School of Physics and Astronomy, University of Southampton,\\[-0em]
Southampton SO17 1BJ, UK \\[2mm]
${}^b$
Instituto de F\'{\i}sica Te\'orica UAM/CSIC,\\[-0em]
Facultad de Ciencias C-XVI, Universidad Aut\'onoma de Madrid, \\[-0em]
Cantoblanco, 28049 Madrid, Spain\\[2mm]
}
\small{\bf Abstract} \\[3mm]
\end{center}
{\small
We perform a complete classification of the flux-induced 12d algebras  compatible with the set of ${\cal N}=1$ type II orientifold models that are T-duality invariant, and allowed by the symmetries of the $\,\mathbb{T}^{6}/ (\mathbb{Z}_{2} \times \mathbb{Z}_{2})\,$ isotropic orbifold. The classification is performed in a type IIB frame, where only $\bar{H}_3$ and $Q$ fluxes are present. We then study no-go theorems, formulated in a type IIA frame,  on the existence of Minkowski/de Sitter (Mkw/dS) vacua.
By deriving a dictionary between the sources of potential energy in types IIB and IIA, we are able to combine algebra results and no-go theorems. The outcome is a systematic procedure for identifying  phenomenologically viable models where Mkw/dS vacua may exist. 
We present a complete table of the allowed algebras and the viability of their resulting scalar potential, and we point at the models which stand any chance of producing a fully stable vacuum.
}
\vspace{40mm}
\begin{flushleft}
\rule{170mm}{0.5pt}\\
\begin{footnotesize}e-mail: 
b.de-carlos@soton.ac.uk , adolfo.guarino@uam.es , jesus.moreno@uam.es 
\end{footnotesize} \\
\end{flushleft}

\newpage
\setcounter{page}{1}
\pagestyle{plain}
\renewcommand{\thefootnote}{\arabic{footnote}}
\setcounter{footnote}{0}

\tableofcontents

\section{Motivation and outline}
\label{sec:intro}

The ten dimensional Supergravities arising as the low energy limit of the different string theories are found to be related by a set of duality symmetries. One of such symmetries that has been deeply studied in the literature is T-duality, which relates  type IIA and type IIB  Supergravities. After reducing these theories from ten to four dimensions, including flux backgrounds for the universal NS-NS $\,3$-form $\,H_{3}$ and the set of R-R $\,p$-forms $\,F_{p}\,$, T-duality is no longer present at the level of the 4d effective models. To restore it, an enlarged set of fluxes known as \textit{generalised fluxes} is required \cite{Shelton:2005cf,Wecht:2007wu}. 

These fluxes are generated by taking a $\,\bar{H}_{3}$ flux and applying a chain of successive T-duality transformations,
\beq
\bar{H}^{}_{abc}  \overset{T_{a}}{\longrightarrow}\omega^{a}_{bc}  \overset{T_{b}}{\longrightarrow}  Q^{ab}_{c}  \overset{T_{c}}{\longrightarrow}  R^{abc}_{}  \ .
\label{Tdualitychain}
\eeq
The first T-duality transformation leads to a new flux associated to the internal components of a spin connection, $\,\omega$. This flux induces a twist on the internal space metric and is called the geometric $\,\omega\,$ flux. Further T-duality transformations give rise to the so-called non-geometric fluxes, $\,Q\,$ and $\,R\,$. 
In their presence, the interpretation of the internal space becomes more subtle.  Only a local, but not global, description is possible when we switch on $\,Q\,$.
And this feature is absent once a $\,R\,$ flux is turned on. Those spaces with a local but not global description are known as T-fold spaces \cite{Hull:2004in,Hull:2006va}. 

The above set of  fluxes determines the Lie algebra $\,\fg\,$ of the Supergravity group $\,\cG$, invariant under T-duality transformations. This algebra is spanned by the isometry  $\,Z_{a}\,$ and  gauge generators $\,X^{a}$, with $\,a=1,\ldots,6\,$, that arise from the reduction of the metric and the $B$ field.  Their commutators %
\beq
\begin{array}{ccrcr}
\left[ Z_{a},Z_{b} \right] &=& \bar{H}_{abc} \, X^{c} &+& \omega_{ab}^{c} \, Z_{c} \ ,\\
\left[ Z_{a}\,,X^{b} \right] &=& -\,\omega^{b}_{ac} \, X^{c} &+& Q_{a}^{bc} \, Z_{c}  \ ,\\
\left[ X^{a},X^{b} \right] &=& Q_{c}^{ab} \, X^{c} &+& R^{abc} \, Z_{c} \ ,
\label{generalalgebra}
\end{array}
\eeq
involve the generalised fluxes, that play the role of structure constants \cite{Shelton:2005cf}. These are subject to several constraints.
Besides the ones imposed by the symmetries of the compactification, they have to fulfil  Jacobi identities and obey the cancellation of the induced tadpoles.

Compactifications of string theory including generalised fluxes have been deeply studied in the literature. In particular, their geometrical properties have been explored in \cite{Hull:2004in,Hull:2006va,Dabholkar:2005ve,Hull:2007jy,Dall'Agata:2007sr,Dall'Agata:2008qz,Grana:2008yw,ReidEdwards:2009nu,Albertsson:2008gq,Samtleben:2008pe,Hull:2005hk,Micu:2007rd}. These compactifications are naturally described as Scherk-Schwarz reductions \cite{SS} on a doubled torus, $\,\mathbb{T}^{12}$, twisted under $\,\cG$. A stringy feature of these reductions is that the coordinates in $\,\mathbb{T}^{12}\,$ account for the ordinary coordinates and their duals, so both momentum and winding modes of the string are treated on equal footing. Furthermore, the fluctuations of the internal components of the metric and the $B$ field are jointly described \cite{Kaloper:1999yr} in terms of a $\,\textrm{O}(6,6)\,$ doubled space metric. In this framework,  a T-duality transformation can be interpreted as a  $\,\textrm{SO}(6,6)\,$ rotation on the background \cite{Dabholkar:2005ve}. 

As far as model building is concerned, these scenarios are very promising. Generalised fluxes induce new terms in the 4d effective potential. As a consequence, mass terms for the moduli may be generated. This mechanism has been largely studied in type IIA string compactifications in the presence of  $\,\bar{H}_{3}\,$,   R-R $\,\bar{F}_{p}\,$  and $\,\omega\,$  geometric  fluxes \cite{Hull:2005hk,Grana:2006kf,Aldazabal:2007sn,Caviezel:2008ik,Derendinger:2004jn,Villadoro:2005cu,Derendinger:2005ph,Camara:2005dc}.  One expects that enlarging the number of fluxes, including the non-geometric ones,  could help providing complete moduli stabilisation \cite{Micu:2007rd,Palti:2007pm}.

We should also point out that, since fluxes are relevant for moduli dynamics, the cosmological implications of those are strongly related to the geometrical properties of the internal space \cite{Roest:2009dq,Hertzberg:2007wc,Caviezel:2008tf,Flauger:2008ad}. For instance, the existence of de Sitter vacua, as required by the observations, needs of a (positive) source of potential energy directly coming from the (negative) curvature of the internal manifold \cite{Silverstein:2007ac,Haque:2008jz, Danielsson:2009ff}.  As we mentioned above, the concept of internal space is distorted or even lost once we include non-geometric fluxes.  Then the interplay between generalised fluxes and moduli stabilisation (or dynamics) has to be decoded from the whole 12d algebra (\ref{generalalgebra}).

In this paper we will work out that interplay in a particular case:  the $\,\mathbb{T}^{6}/ (\mathbb{Z}_{2} \times \mathbb{Z}_{2})\,$ orbifold. To make it more affordable, we will  impose an additional $\,\mathbb{Z}_{3}\,$ symmetry on the fluxes under the exchange of the three tori. We will refer to this restriction as  \textit{isotropic flux background} or, with some abuse of language, \textit{isotropic orbifold}. Notice that this isotropy assumption is realised on the flux backgrounds instead of on the internal space, as it was done in \cite{Shelton:2005cf,Shelton:2006fd}. This fact correlates with the sort of localised sources that can be eventually added \cite{Aldazabal:2006up,Font:2008vd}.
 
 Working in the $\,\mathcal{N}=1\,$ orientifold limit, which allows for O3/O7-planes and forbids the $\,\omega\,$ and  $\,R\,$ fluxes, a classification of all the compatible non-geometric $Q$ flux backgrounds was carried out in ref.~\cite{Font:2008vd}. We go now one step further and extend the results of \cite{Font:2008vd}  to include the $\,\bar{H}_{3}\,$ flux, providing a complete classification of the Supergravity algebras. As  the algebra (\ref{generalalgebra}) is T-duality invariant, this classification does not depend on the choice of orientifold projection. 
 
After completing the classification, we focus our attention on the existence of de Sitter (dS) and Minkowski (Mkw) vacua, which are interesting for phenomenology, i.e. that break Supersymmetry. Some no-go theorems concerning the existence of such vacua have been established, as well as mechanisms to circumvent them \cite{Hertzberg:2007wc,Caviezel:2008tf,Flauger:2008ad,Silverstein:2007ac,Haque:2008jz}. However, they were mostly proposed in the language of a type IIA generalised flux compactification, including O6-planes and D6-branes. We therefore develop a dictionary between the contributions to the scalar potential in the IIA language, in which the no-go theorems were formulated, and the IIB one in which we performed the classification of the Supergravity algebras. It is a IIA $\leftrightarrow$ IIB mapping between both effective model descriptions. By means of this dictionary, we exclude the existence of dS/Mkw vacua in more than {\bf half} of the effective models based on non-semisimple Supergravity algebras. On the other hand, those based on semisimple algebras survive the no-go theorem and stand a chance of having all moduli stabilised.

With the set of effective models that are phenomenologically interesting (aka SUSY breaking ones) narrowed down to a few, a detailed numerical study of potential vacua will be presented in a forthcoming paper \cite{deCarlos:2009qm}.

\section{Fluxes and Supergravity algebras}
\label{sec:algebra}

In the absence of fluxes, compactifications of the type II ten dimensional Supergravities on $\,\mathbb{T}^{6}\,$ orientifolds yield a $\,\mathcal{N}=4\,$, $\,d=4\,$ Supergravity. Without considering additional vector multiplets coming from D-branes, its deformations produce $\,\mathcal{N}=4\,$ gauged Supergravities \cite{Schon:2006kz} specified by two constant embedding tensors, $\,\xi_{\alpha A}\,$ and $\,f_{\alpha ABC}\,$, under the global symmetry
\beq
\textrm{SL}(2,\mathbb{Z}) \times \textrm{SO}(6,6,\mathbb{Z})\ , 
\label{global}
\eeq
where $\,\alpha=\pm\,$ and $\,A,B,C=1,\ldots,12\,$. These embedding tensors are interpreted as flux parameters, so the fluxes become the gaugings of the $\,\mathcal{N}=4\,$ gauged Supergravity \cite{Samtleben:2008pe}.

In this work we focus on the orientifold limits of the $\,\mathbb{T}^{6}/(\mathbb{Z}_{2} \times \mathbb{Z}_{2})\,$ orbifold for which the global symmetry (\ref{global}) is broken to the $\,\textrm{SL}(2,\mathbb{Z})^{7}\,$ group and the tensor $\,\xi_{\alpha A}\,$ is projected out. Compactifying the type II Supergravities on this orbifold produces a $\,\mathcal{N}=2\,$ Supergravity further broken to $\,\mathcal{N}=1\,$ in its orientifold limits. 

\subsection{The $\,\mathcal{N}=1\,$ orientifold limits as duality frames}
\label{sec:N=1}

The $\,\mathcal{N}=1\,$ effective models based on type II orientifold limits of toroidal orbifolds allow for localised objects of negative tension, known as O$p$-planes, located at the fixed points of the orientifold involution action. These orientifold limits are related by a chain of T-duality transformations \cite{Aldazabal:2006up},
\beq
\textrm{type IIB with O3/O7} \,\, \leftrightarrow \,\,\textrm{type IIA with O6} \,\,\leftrightarrow\,\,\textrm{type IIB with O9/O5} \ ,
\eeq
so we will often refer to them as \textit{duality frames}.

Each of these frames projects out half of the flux entries. The IIB orientifold limit allowing for O3/O7-planes projects the geometric $\,\omega\,$ and the non-geometric $\,R\,$ fluxes out of the effective theory. This duality frame is particularly suitable when classifying the Supergravity algebras, since it does not forbid certain components in all the fluxes, as it happens with the IIA orientifold limit allowing for O6-planes, but certain fluxes as a whole\footnote{This is also the case for the IIB orientifold limit allowing for O9/O5-planes, which forbids the $\,\bar{H}_{3}\,$ and $\,Q\,$ fluxes. The generalised fluxes mapping between the O3/O7 and O9/O5 orientifold limits reads $\,Q^{ab}_{c} \leftrightarrow \omega^{c}_{ab}\,$ together with $\,\bar{H}_{abc} \leftrightarrow R^{abc}$.}. In this duality frame, the Supergravity algebra (\ref{generalalgebra}) simplifies to
\beq
\begin{array}{ccr}
\left[ X^{a},X^{b} \right] &=& Q_{c}^{ab} \, X^{c}  \ ,\\
\left[ Z_{a}\,,X^{b} \right] &=& Q_{a}^{bc} \, Z_{c} \, \ , \\
\left[ Z_{a},Z_{b} \right] &=& \bar{H}_{abc} \, X^{c}  \ ,
\label{IIBalgebra}
\end{array}
\eeq
and the effective models admit a description in terms of a reduction on a T-fold space. From now on, we will refer to the IIB orientifold limit allowing for O3/O7-planes as the \textit{T-fold description} of the effective models. One observes that (\ref{IIBalgebra}) comes up with a gauge-isometry $\,\mathbb{Z}_{2}$-graded structure involving the subspaces expanded by the gauge $\,X^{a}\,$ and the isometry $\,Z_{a}\,$ generators as the grading subspaces.

In the T-fold description, the Supergravity group $\,\cG\,$  has a six dimensional subgroup $\,\cG_{gauge}\,$ whose algebra $\,\fg_{gauge}\,$ involves the vector fields $\,X^{a}\,$ coming from the reduction of the $B$-field. $\,\fg_{gauge}\,$ is completely determined by the non-geometric $\,Q\,$ flux, forced to satisfy the quadratic $XXX$-type Jacobi identity $\,Q^{2}=0\,$ from (\ref{IIBalgebra}),
\beq
Q^{[ab}_{x}\,Q_{d}^{c]x} = 0 \ .
\label{QQJacobi}
\eeq
From the general structure of (\ref{IIBalgebra}), the remaining $\,Z_{a}\,$ vector fields coming from the reduction of the metric are the generators of the reductive and symmetric coset space $\,\cG/\cG_{gauge}\,$ \cite{Manousselis:2001re}. Provided a $\,Q\,$ flux, the mixed gauge-isometry brackets in (\ref{IIBalgebra}) are given by the co-adjoint action $\,Q^{*}\,$ of $\,Q\,$ and the $\,\cG/\cG_{gauge}$ coset space is determined by the $\,\bar{H}_{3}\,$ flux restricted by the $\,\bar{H}_{3}Q=0\,$ constraint
\beq
\bar{H}_{x[bc}\,Q_{d]}^{ax} = 0 \ ,
\label{QHJacobi}
\eeq
coming from the quadratic $XZZ$-type Jacobi identity from (\ref{IIBalgebra}). Any point in the coset space remains fixed under the action of the isotropy subgroup $\,\cG_{gauge}\,$ of $\,\cG\,$ \cite{Castellani:1999fz}, so an effective model is defined by specifying both the Supergravity algebra $\,\fg\,$ as well as the subalgebra $\,\fg_{gauge}\,$ associated to the isotropy subgroup of the coset space $\,\cG/\cG_{gauge}$.

\subsection{T-dual algebras in the isotropic $\,\mathbb{Z}_{2} \times \mathbb{Z}_{2}\,$ orientifolds}
\label{sec:algebras}

The fluxes needed to make the 4d effective models based on the $\,\mathbb{Z}_{2} \times \mathbb{Z}_{2}\,$ orbifold invariant under $\,\textrm{SL}(2,\mathbb{Z})$ modular transformations on each of its seven untwisted moduli were introduced in \cite{Aldazabal:2006up}. Furthermore, the set of $\,\textrm{SL}(2,\mathbb{Z})^{7}$-invariant isotropic flux backgrounds consistent with $\,\mathcal{N}=1\,$ Supersymmetry was found to be systematically computable \cite{Guarino:2008ik} from the set of T-duality invariant ones previously derived in \cite{Font:2008vd}. However an exhaustive identification of the Supergravity algebras underlying such T-duality invariant isotropic flux backgrounds remains undone, and that is what we present in this section.  Since $\,\fg\,$ is invariant under T-duality transformations, this classification of algebras is valid in any duality frame although we are computing it in the IIB orientifold limit allowing for O3/O7-planes.

An exploration of their $\,\mathcal{N}=4\,$ origin, if any, after removing the orbifold projection, is beyond the scope of this work. Nevertheless, recent progress on this bottom-up approach has been made for the set of geometric type IIA flux compactifications \cite{Dall'Agata:2009gv}, complementing the previous work \cite{Aldazabal:2008zza} that focused on non-geometric type IIB flux compactifications.

\subsubsection{The set of gauge subalgebras}

The discrete $ \, \mathbb{Z}_{2} \times \mathbb{Z}_{2} \, $ orbifold symmetry together with the cyclic $\,\mathbb{Z}_{3}\,$ symmetry (isotropy) of the fluxes under the exchange $\,1\rightarrow 2\rightarrow 3\,$ in the factorisation
\beq
\mathbb{T}^{6} = \mathbb{T}_{1}^{2} \times  \mathbb{T}_{2}^{2} \times \mathbb{T}_{3}^{2} \ ,
\label{TorusFact}
\eeq
select the simple $\,\mathfrak{so(3)} \sim \mathfrak{su(2)}\,$ algebra \cite{Derendinger:2004jn} as the fundamental block for building the set of compatible $\,\fg_{gauge}\,$ subalgebras within the $\,\mathcal{N}=1\,$ algebra (\ref{IIBalgebra}). 

The two maximal $\,\fg_{gauge}\,$ subalgebras that our orbifold admits are the semisimple $\, \mathfrak{so(4)} \sim \mathfrak{su(2)^{2}}\,$ and $\, \mathfrak{so(3,1)}\,$ Lie algebras. Both possibilities come up with a $\mathbb{Z}_{2}$-graded structure differing in the way in which the two $\,\mathfrak{su(2)}\,$ factors are glued together when it comes to realising the grading. 

Since there is no additional restriction over $\,\fg_{gauge}\,$, apart from that of respecting the isotropic orbifold symmetries, any $\,\mathbb{Z}_{2}$-graded contraction\footnote{We refer the reader interested in the topic of G-graded Lie algebras and their contractions to refs~\cite{deMontigny:1990cw,weimar-woods}.} of the previous maximal subalgebras is also a valid $\,\fg_{gauge}$. The set of such contractions comprises the non-semisimple subalgebras of $\,\,\mathfrak{iso(3)} \sim \mathfrak{su(2)} \oplus_{\mathbb{Z}_{3}} \mathfrak{u(1)^{3}}\,\,$ and $\,\,\mathfrak{nil} \sim \mathfrak{u(1)^{3}} \oplus_{\mathbb{Z}_{3}} \mathfrak{u(1)^{3}}\,\,$ arising from continuous contractions (where $\,\mathfrak{nil} \equiv n\,3.5\,$ in \cite{Grana:2006kf}), together with the direct sum $\,\mathfrak{su(2)} + \mathfrak{u(1)^{3}}\,$ coming from a discrete contraction \cite{deMontigny:1990cw}. The $\,\oplus_{\mathbb{Z}_{3}}\,$ symbol denotes the semidirect sum of algebras endowed with the $\,\mathbb{Z}_{3}\,$ cyclic structure coming from isotropy. These subalgebras were already identified in \cite{Font:2008vd}.

Denoting $\, ( E^{I}, \wt E^{I} )_{I = 1,2,3}\,$ a basis for $\,\fg_{gauge}\,$, the entire set of gauge subalgebras previously found is gathered in the brackets 
\beq
[ E^{I} , E^{J}  ]  =  \kappa_{1} \,   \eps_{IJK} \, E^{K}  \hspace{5mm} , \hspace{5mm} [ E^{I} , \wt E^{J}  ]  =  \kappa_{12} \,   \eps_{IJK} \, \wt E^{K} \hspace{5mm} , \hspace{5mm} [ \wt E^{I} ,  \wt E^{J}  ]  = \kappa_{2} \,  \eps_{IJK} \, E^{K}  \ ,
\label{EGen}
\eeq
with an antisymmetric $\,\epsilon_{IJK}\,$ structure imposed by the isotropy $\,\mathbb{Z}_{3}\,$ symmetry. The structure constants $\,\mathcal{Q}\,$ given by
\beq
\mathcal{Q}_{E^{I},E^{J}}^{E^{K}}=\kappa_{1} \hspace{10mm},\hspace{10mm}  \mathcal{Q}_{E^{I}, \wt E^{J}}^{\wt E^{K}}=\kappa_{12} \hspace{10mm} \textrm{and} \hspace{10mm} \mathcal{Q}_{\wt E^{I},\wt E^{J}}^{E^{K}}=\kappa_{2} \ ,
\label{Qcanonic}
\eeq
are restricted by the Jacobi identities $\,\mathcal{Q}^{2}=0\,$ to either 
\beq
\kappa_{1}=\kappa_{12} \hspace{1cm} \,\,\, \mbox{ or } \,\,\, \hspace{1cm} \kappa_{12}=\kappa_{2}=0   \ .
\label{kappaJacobi}
\eeq
The first solution in (\ref{kappaJacobi}) gives rise to the maximal gauge subalgebras and their continuous contractions, whereas the second generates the discrete contraction. The intersection between both spaces of solutions contains just the trivial point $\,\kappa_{1}=\kappa_{12}=\kappa_{2}=0\,$. The structure constants in (\ref{Qcanonic}) can always be normalised to $\,1$, $0$ or $-1\,$ by a rescaling of the generators in (\ref{EGen}). These normalised $\kappa$-parameters are presented in table~\ref{tablekappa}. 
\begin{table}[htb]
\small{
\renewcommand{\arraystretch}{1.25}
\begin{center}
\begin{tabular}{|c|c|c|c|c|c|c|}
\hline
$\fg_{gauge}$ & $\mathfrak{so(3,1)}$ & $\mathfrak{so(4)}$ &  $\mathfrak{iso(3)}$ & $\mathfrak{nil}$ & $\mathfrak{su(2)+u(1)^3}$ &$ \mathfrak{u(1)^{6}}$  \\
\hline
$\kappa_{1}$ & $1$  & $1$  & $1$ & $0$ & $1$ & $0$  \\
\hline
$\kappa_{12}$  & $1$  & $1$  & $1$ & $0$ & $0$ & $0$ \\
\hline
$\kappa_{2}$ & $-1$  & $1$  & $0$ & $1$ & $0$ & $0$ \\
\hline
\end{tabular}
\end{center}
\caption{The set of normalised gauge subalgebras satisfying ($2.9$).}
\label{tablekappa}
}
\end{table}

In the following, we will refer to the $\, ( E^{I} , \wt E^{I} ) \,$ generator basis equipped with the $\,\mathcal{Q}\,$ structure constants of (\ref{Qcanonic}), as the \textit{canonical basis} for $\,\fg_{gauge}$.

\subsubsection{The extension to a full Supergravity algebra}

Thus far, we have explored the set of $\,\fg_{gauge}\,$ compatible with the isotropic $\mathbb{Z}_{2} \times \mathbb{Z}_{2}$ orbifold symmetries finding that there exists a gauge $\,\mathbb{Z}_{2}$-graded inner structure modding out all of them. Specifically, this set consists of the two maximal semisimple $\,\fg_{gauge}$, those of $\,\mathfrak{so(3,1)}\,$ and $\,\mathfrak{so(4)}$, and their non-semisimple $\,\mathbb{Z}_{2}$-graded contractions. 
\\[2mm]
Two questions that arise at this point are the following
\begin{enumerate}
\item How does $\,\fg_{gauge}\,$ in (\ref{EGen}) extend to a twelve dimensional Supergravity algebra $\,\mathfrak{g}\,$? Since we are dealing with an orientifold limit of the isotropic $\,\mathbb{Z}_{2} \times \mathbb{Z}_{2}\,$ orbifold, the structure constants of $\,\mathfrak{g}\,$ can be classified according to the group $\,\textrm{SO}(2,2) \times \textrm{SO}(3) \subset \textrm{SO}(6,6)\,$ with the embedding $\,(\textbf{4},\textbf{3})=\textbf{12}\,$ \cite{Derendinger:2004jn}. The $\,\textrm{SO}(3)\,$ factor accounts for the cyclic $\,\mathbb{Z}_{3}\,$ isotropy symmetry and imposes a $\,\eps_{IJK}\,$ structure, not only in the gauge brackets of (\ref{EGen}), but also in the extended brackets involving the isometry generators. The $\,\textrm{SO}(2,2)\,$ factor reflects on a splitting of $\,\mathfrak{g}\,$ into four $\,( E, \wt E , D , \wt D )\,$ algebra subspaces expanded by the gauge generators $\,( E^{I} , \wt E^{I} )_{I=1,2,3}\,$ of (\ref{EGen}) and a new set of isometry generators $\,( D_{I} , \wt D_{I} )_{I=1,2,3}\,$. 
In addition to the gauge brackets specified by $\,\mathcal{Q}\,$ in (\ref{Qcanonic}), the algebra $\,\mathfrak{g}\,$ will involve an enlarged set of structure constants 
\beq
\mathcal{Q^{*}}_{(E^{I},\wt E^{I})\,,\,(D_{J},\wt D_{J})}^{(D_{K},\wt D_{K})} \hspace{10mm} \textrm{and} \hspace{10mm} \mathcal{H}_{(D_{I},\wt D_{I})\,,\,(D_{J},\wt D_{J})}^{(E^{K},\wt E^{K})}\ ,
\label{Q*Hcanonic}
\\[2mm]
\eeq
such that the mixed gauge-isometry brackets in (\ref{Q*Hcanonic}) are given by the co-adjoint action $\,\mathcal{Q}^{*}$ of $\,\mathcal{Q}\,\,$ and $\,\,( D_{I} , \wt D_{I} )_{I=1,2,3}\,$ become the generators of the reductive and symmetric coset space $\,\cG/\cG_{gauge}\,$ \cite{Manousselis:2001re}.

\item Does such an extension result in a $\,G$-graded structure? If it does, the $\,G$-grading of $\,\fg\,$ has to accommodate for both gauge-inner and gauge-isometry $\,\mathbb{Z}_{2}$-graded structures of (\ref{Qcanonic}) and (\ref{Q*Hcanonic}). This reduces the candidates to the $\,G=\mathbb{Z}_{2} \oplus \mathbb{Z}_{2}\,$, $\,G=\mathbb{Z}_{2} \otimes \mathbb{Z}_{2}\,$ and $\,G=\mathbb{Z}_{4}\,$ groups. 
\end{enumerate}

Let us start by deriving the extension of the $\,\fg_{gauge}\,$ based on the $\,\kappa_{1}=\kappa_{12}\,$ solution in (\ref{kappaJacobi}) to a full Supergravity algebra $\,\fg$. For the extended Jacobi identity $\,\mathcal{H}\mathcal{Q}=0\,$ to be fulfilled, the most general twelve dimensional Supergravity algebra is given (up to redefinitions of the algebra  basis) in table~\ref{kappacontsol}, where the $\,( \epsilon_{1} , \epsilon_{2} )\,$ real quantities determine the new entries in the extended structure constants $\,\mathcal{H}\,$, involving the brackets between the isometry generators. Therefore, the $\epsilon$-parameters determine the coset space $\,\cG/\cG_{gauge}\,$ and the Supergravity algebra $\,\fg \,$ built from a specific $\,\fg_{gauge}\,$ \cite{Castellani:1999fz}. Observe that $\,\fg\,$ has a non manifest graded structure. Although we omit the tedious proof, it can always be transformed into a $\,G$-graded form with $\,G=\mathbb{Z}_{2} \oplus \mathbb{Z}_{2}\,$, $\,G=\mathbb{Z}_{2} \otimes \mathbb{Z}_{2}\,$ and $\,G=\mathbb{Z}_{4}\,$, by an appropriate rotation of the $\,(E,\wt E, D,\wt D)\,$ vector of $\,\textrm{SO}(2,2)\,$ without mixing the gauge and isometry
subspaces\footnote{$\, \mathbb{Z}_{2} \otimes \mathbb{Z}_{2}\,$ acts naturally on the bi-complex numbers. The maximal Supergravity algebra with this  grading is  $\,\fg=\mathfrak{so(3,1)^{2}}$, the $\,\mathfrak{su(2)}$ bicomplexification.}. 

\begin{table}[htb]
\renewcommand{\arraystretch}{1.30}
\begin{center}
\begin{tabular}{|c||c|c|c|c|}
\hline
$\kappa_{1}=\kappa_{12}$              & $E^{J}$         & $\wt E^{J}$   & $D_{J}$               & $\wt D_{J}$  \\
\hline
\hline
$E^{I}$        & $\kappa_{1} \, E^{K}$        & $\kappa_{1} \, \wt E^{K}$  &      $\kappa_{1} \,  D_{K}$           &  $\kappa_{1} \, \wt D_{K}$   \\
\hline
$\wt E^{I}$ & $\kappa_{1} \, \wt E^{K}$  & $\kappa_{2} \, E^{K}$      & $- \kappa_{2} \, \wt D_{K}$    &  $- \kappa_{1} \, D_{K}$ \\
\hline
$D_{I} $      &  $\kappa_{1} \, D_{K}$        & $- \kappa_{2} \, \wt D_{K}$  &$ -\epsilon_{1} \, \kappa_{2} \, E^{K}  - \epsilon_{2} \, \kappa_{2} \, \wt E^{K}$  &   $\epsilon_{2} \, \kappa_{2} \, E^{K} + \epsilon_{1} \, \kappa_{1} \, \wt E^{K}$ \\
\hline
$\wt D_{I}$ & $\kappa_{1} \, \wt D_{K}$ &  $- \kappa_{1} \, D_{K} $     &  $\epsilon_{2} \, \kappa_{2} \, E^{K}  + \epsilon_{1} \, \kappa_{1} \,\wt E^{K}$ &  $-\epsilon_{1} \, \kappa_{1} \,  E^{K} - \epsilon_{2} \, \kappa_{1} \, \wt E^{K}$ \\
\hline
\end{tabular}
\end{center}
\caption{Commutation relations for the Supergravity algebra $\,\fg\,$ based on the $\,\kappa_{1}=\kappa_{12}\,$ solution in (2.9).}
\label{kappacontsol}
\end{table}

Working out the extension of the $\,\fg_{gauge}\,$ for $\,\kappa_{12}=\kappa_{2}=0\,$ solution in (\ref{kappaJacobi}), the most general Supergravity algebra verifying $\,\mathcal{H}\mathcal{Q}=0\,$ is written (again up to redefinitions of the algebra  basis) in table~\ref{kappadiscsol}, resulting in an explicit $\,G=\mathbb{Z}_{2} \oplus \mathbb{Z}_{2}\,$ graded structure. It factorizes into the direct sum of two six dimensional pieces spanned by $\,(E^{I},D_{I})\,$ and $\,(\wt E^{I},\wt D_{I})\,$ respectively. We will refer to the $\, ( E^{I} , \wt E^{I} \,;\, D_{I} , \wt D_{I}  )_{I=1,2,3} \,$ generators, satisfying the commutation relations either in table~\ref{kappacontsol} or \ref{kappadiscsol}, as the \textit{canonical basis} of $\,\fg$. The structure constants $\,\mathcal{Q}\,$ and $\,\mathcal{H}\,$ in this basis depend on the $\,(\kappa_{1},\kappa_{2},\eps_{1},\eps_{2})\,$ parameters and can be directly read from there.

\begin{table}[htb]
\renewcommand{\arraystretch}{1.30}
\begin{center}
\begin{tabular}{|c||c|c|c|c|}
\hline
$\kappa_{12}=\kappa_{2}=0$      & $E^{J}$  & $\wt E^{J}$   &  $D_{J}$         &  $\wt D_{J} $       \\
\hline
\hline
$E^{I}$        & $\kappa_{1} \, E^{K}$   & $0$  &  $\kappa_{1} \,  D_{K} $                & $0$   \\
\hline
$\wt E^{I} $ & $0$    & $0$ &  $0$      & $0$    \\
\hline
$D_{I} $    & $\kappa_{1} \, D_{K} $  & $0$   & $-\eps_{1} \, \kappa_{1} \, E^{K}  $    &  $0$  \\
\hline
$\wt D_I $  & $0$  & $0$ &  $0$   &  $ -\eps_{2} \, \kappa_{1} \, \wt E^{K} $    \\
\hline
\end{tabular}
\end{center}
\caption{Commutation relations for the Supergravity algebra $\,\fg\,$ based on the $\,\kappa_{12}=\kappa_{2}=0\,$ solution in (2.9).}
\label{kappadiscsol}
\end{table}

A powerful clue to identifying the set of Supergravity algebras that can be realized within the brackets in tables~\ref{kappacontsol} and \ref{kappadiscsol} comes from the study of their associated Killing-Cartan matrix, denoted $\,\mathcal{M}\,$. It has a block-diagonal structure
\beq
\mathcal{M} = \textrm{Diag}\left( \mathcal{M}_{g} , \mathcal{M}_{g} , \mathcal{M}_{g} , \mathcal{M}_{isom} , \mathcal{M}_{isom} , \mathcal{M}_{isom}  \right) \ ,
\eeq
where $\,\mathcal{M}_{g}\,$ and $\,\mathcal{M}_{isom}\,$ are $\,2 \times 2\,$ matrices referring to the pairs $\,( E^{I} , \wt E^{I} )\,$ and $\,( D_{I} , \wt D_{I} )\,$ of generator subspaces, respectively. Let us study the diagonalisation of $\,\mathcal{M}$. The two eigenvalues of the $\,\mathcal{M}_g\,$ matrix
\beq
\begin{array}{rclcl}
\kappa_{1}=\kappa_{12} :   &  & \lambda^{(1)}_{gauge} = -2^{3} \, \kappa_{1}^{2}     & \hspace{5mm} \mbox{ and } \hspace{5mm} & \hspace{5mm}\lambda^{(2)}_{gauge} =  -2^{3} \, \kappa_{1} \, \kappa_{2}    \ , \\[2mm]
\kappa_{12}=\kappa_{2}=0 :  &  & \lambda^{(1)}_{gauge} =  -2^{2} \, \kappa_{1}^{2}   & \hspace{5mm} \mbox{ and } \hspace{5mm} & \hspace{5mm}\lambda^{(2)}_{gauge} = 0 \ ,
\end{array}
\label{Eigenvalues}
\eeq
are obtained by substituting the normalised $\,\kappa$-configurations in table~\ref{tablekappa}. For the $\,\mathcal{M}_{isom}\,$ matrix, they are computed by solving the characteristic polynomial
\beq
\lambda_{isom}^{2} - \textrm{T} \, \lambda_{isom} + \textrm{D} = 0 \ ,
\label{characpol}
\eeq
determined by its trace T and its determinant D. Those are given by
\beq
\begin{array}{rclcl}
\kappa_{1}=\kappa_{12} :   &  & \textrm{T}=2^{3} \, \eps_{1} \, \kappa_{1} \, (\kappa_{1} + \kappa_{2})      & \hspace{5mm} \mbox{ and } \hspace{5mm} & \hspace{5mm} \textrm{D} = 2^{6} \, \kappa_{1}^{2} \, \kappa_{2} \, ( \eps_{1}^{2} \, \kappa_{1} - \eps_{2}^{2} \, \kappa_{2} )    \ , \\[2mm]
\kappa_{12}=\kappa_{2}=0 :  &  & \textrm{T} = 12 \, \eps_{1} \, \kappa_{1}^{2}   & \hspace{5mm} \mbox{ and } \hspace{5mm} & \hspace{5mm}\textrm{D} = 0 \ .
\end{array}
\label{TandD}
\eeq
Provided the $\,\kappa_{1}=\kappa_{12}\,$ solution in (\ref{kappaJacobi}), the Supergravity algebra $\,\fg\,$ becomes semisimple iff
\beq
\kappa_{1}\,\kappa_{2} \, (\kappa_{1} \, \eps_{1}^{2} - \kappa_{2} \, \eps_{2}^{2}) \neq 0\ , 
\label{semisimplecond}
\eeq
whereas for the $\,\kappa_{12}=\kappa_{2}=0\,$ solution $\,\fg\,$ is always non-semisimple since a null $\,\lambda_{isom}=0\,$ eigenvalue comes out of (\ref{characpol}). 

Some of the generators in the adjoint representation of $\,\fg\,$ may vanish when we are dealing with a non-semisimple algebra. If so, this representation is no longer faithful and the Supergravity algebra $\,\fg_{emb}\,$ realised on the curvatures and embeddable within the $\,\mathfrak{o(6,6)}\,$ duality algebra becomes smaller than the algebra $\,\fg\,$ involving the vector fields \cite{Dall'Agata:2007sr}. 

After a detailed exploration, the set of $\,\mathfrak{g}\,$ allowed by the $\,\mathcal{N}=1\,$ orientifolds of the isotropic $\,\mathbb{Z}_{2} \times \mathbb{Z}_{2}\,$ orbifold are listed in table~\ref{tablelistalgebras}. The spectrum includes from the flux-vanishing $\,\fg=\mathfrak{u(1)^{12}}\,$ case up to the most involved $\,\fg=\mathfrak{so(3,1)^{2}}\,$ algebra. All of them are $\,G$-graded contractions of those Supergravity algebras built from the maximal $\,\fg_{gauge}=\mathfrak{so(4)}\,$ and $\,\fg_{gauge}=\mathfrak{so(3,1)}\,$ subalgebras. Specifically, contractions based on the abelian $\,G=\mathbb{Z}_{2} \oplus \mathbb{Z}_{2}\,$, $\,G=\mathbb{Z}_{2} \otimes \mathbb{Z}_{2}\,$ and $\,G=\mathbb{Z}_{4}\,$ finite groups, compatible with the isotropic orbifold symmetries.

\begin{table}
\small{
\renewcommand{\arraystretch}{1.50}
\begin{center}
\begin{tabular}{|c|c|c|c|c|c|}
\hline 
$\#$  & $\,\fg_{gauge}\,$ & $ \fg $ & $ \fg_{emb} $ & \multicolumn{2}{c|}{ $\,\mathcal{H}\,$ \textsc{extension} } \\
\hline
\hline
%
$ 1 $ & \multirow{2}{*}{ $\mathfrak{so(3,1)}$} &$\mathfrak{so(3,1)^{2}}$ & \multirow{2}{*}{$\fg$} & \multicolumn{2}{c|}{$ \eps_{1}^{2} + \eps_{2}^{2} \neq 0$}\\
\cline{1-1}\cline{3-3} \cline{5-6}$ 2 $ & & $\mathfrak{so(3,1)}   \oplus_{\mathbb{Z}_{3}}  \mathfrak{u(1)^6}$ &  & \multicolumn{2}{c|}{$ \eps_{1}^{2} + \eps_{2}^{2}=0$ }\\
\hline
\hline
%
%
$ 3 $ & \multirow{6}{*}{ $ \mathfrak{so(4)}$} & $\mathfrak{so(3,1)^2}$ & \multirow{6}{*}{ $\fg$} & \multicolumn{2}{c|}{$(\eps_{1}+\eps_{2})>0 \,\,\,\textrm{ , }\,\,\, (\eps_{1} - \eps_{2}) > 0$} \\
\cline{1-1}\cline{3-3} \cline{5-6} $ 4 $ & & $\mathfrak{iso(3)^2}$ &  & \multicolumn{2}{c|}{$(\eps_{1}+\eps_{2})=0 \,\,\,\textrm{ , }\,\,\, (\eps_{1} - \eps_{2}) = 0$}  \\
\cline{1-1}\cline{3-3} \cline{5-6} $ 5 $ & & $\mathfrak{so(4)^2}$ &  & \multicolumn{2}{c|}{$(\eps_{1}+\eps_{2})<0 \,\,\,\textrm{ , }\,\,\, (\eps_{1} - \eps_{2}) < 0$}  \\
\cline{1-1}\cline{3-3} \cline{5-6} $ 6 $ & & $\mathfrak{so(3,1)} + \mathfrak{iso(3)}$ &  & \multicolumn{2}{c|}{$(\eps_{1}+\eps_{2})\geqslant 0 \,\,\,\textrm{ , }\,\,\, (\eps_{1} - \eps_{2}) \eqslantgtr 0$}  \\
\cline{1-1}\cline{3-3} \cline{5-6} $ 7 $ & & $\mathfrak{so(3,1)} + \mathfrak{so(4)}$ &  & \multicolumn{2}{c|}{$(\eps_{1}+\eps_{2}) \gtrless 0 \,\,\,\textrm{ , }\,\,\, (\eps_{1} - \eps_{2}) \lessgtr 0$}  \\
\cline{1-1}\cline{3-3} \cline{5-6} $ 8 $ & & $\mathfrak{iso(3)} + \mathfrak{so(4)}$ &  & \multicolumn{2}{c|}{$(\eps_{1}+\eps_{2}) \eqslantless 0 \,\,\,\textrm{ , }\,\,\, (\eps_{1} - \eps_{2}) \leqslant 0$}  \\
\hline
\hline

%
%
$9$ &\multirow{3}{*}{ $ \mathfrak{iso(3)}$} & $\mathfrak{so(3,1)} \,\oplus_{{\mathbb{Z}_{3}}}\, \mathfrak{u(1)^{6}}$ & \multirow{3}{*}{$\fg$} & $\eps_{1} > 0$ & \multirow{3}{*}{$\eps_{2} = $ \emph{free}} \\
\cline{1-1}\cline{3-3} \cline{5-5} $10$ & & $\mathfrak{iso(3)} \,\oplus_{{\mathbb{Z}_{3}}}\, \mathfrak{u(1)^{6}}$ &  & $\eps_{1} = 0$ &   \\
\cline{1-1}\cline{3-3} \cline{5-5} $11$ & & $\mathfrak{so(4)} \,\oplus_{{\mathbb{Z}_{3}}}\, \mathfrak{u(1)^{6}}$ &  & $\eps_{1} < 0$ &   \\
\hline
\hline
$ 12 $ & \multirow{2}{*}{ $ \mathfrak{nil}$} & $\,\mathfrak{nil}_{12}\mathfrak{(4)}\,$ &  $\,\mathfrak{nil}_{12}\mathfrak{(3)}\,$ & \multirow{2}{*}{$\, \eps_{1}=$ \emph{free}} & $\eps_{2} \neq 0$  \\
\cline{1-1}\cline{3-4}\cline{6-6} $ 13 $ & & $\,\mathfrak{nil}_{12}\mathfrak{(2)}\,$ & $\mathfrak{u(1)^{12}}$ &  & $\eps_{2} = 0$ \\
\hline
\hline
%
%
$ 14 $ & \multirow{6}{*}{ $ \mathfrak{su(2)} + \mathfrak{u(1)^{3}}$} & $\mathfrak{so(3,1)} + \mathfrak{nil}$ & \multirow{2}{*}{ $ \mathfrak{so(3,1)} + \mathfrak{u(1)^{6}}$} & \multirow{2}{*}{$\eps_{1} > 0$} & $\eps_{2} \neq 0$ \\
\cline{1-1}\cline{3-3} \cline{6-6} $ 15 $ & & $\mathfrak{so(3,1)} + \mathfrak{u(1)^{6}}$ &  &  & $\eps_{2} = 0$  \\
\cline{1-1}\cline{3-6} $ 16 $ &  & $\mathfrak{iso(3)} + \mathfrak{nil}$ & \multirow{2}{*}{ $ \mathfrak{iso(3)} + \mathfrak{u(1)^{6}}$} & \multirow{2}{*}{$\eps_{1} = 0$} & $\eps_{2} \neq 0$ \\
\cline{1-1}\cline{3-3} \cline{6-6} $ 17 $ &  & $\mathfrak{iso(3)} + \mathfrak{u(1)^{6}}$ &  &  & $\eps_{2} = 0$  \\
\cline{1-1}\cline{3-6}  $ 18 $ & & $\mathfrak{so(4)} + \mathfrak{nil}$ & \multirow{2}{*}{ $ \mathfrak{so(4)} + \mathfrak{u(1)^{6}}$} & \multirow{2}{*}{$\eps_{1} < 0$} & $\eps_{2} \neq 0$ \\
\cline{1-1}\cline{3-3} \cline{6-6} $ 19 $ & & $\mathfrak{so(4)} + \mathfrak{u(1)^{6}}$ &  &  & $\eps_{2} = 0$  \\
\hline
\hline
%
%
$ 20 $ & $ \mathfrak{u(1)^{6}}$ & $\,\mathfrak{nil}_{12}\mathfrak{(2)}\,$ &  $\,\mathfrak{u(1)^{12}}\,$ & \multicolumn{2}{c|}{ \textsc{unconstrained} }\\
\hline
\end{tabular}
\end{center}
\caption{List of the $\,\mathcal{N}=1\,$ Supergravity algebras $\,\fg\,$ allowed by the isotropic $\,\mathbb{Z}_{2} \times \mathbb{Z}_{2}\,$ orbifold. The $\,\kappa$-parameters fixing $\,\fg_{gauge}\,$ are taken to their normalised values shown in table~$1$. The number $\,p\,$ within the parenthesis in the twelve dimensional $\,\mathfrak{nil}_{12}(p)\,$ nilpotent algebras denotes the nilpotency order. }
\label{tablelistalgebras}
}
\end{table}

It can be observed that $\,\fg=\mathfrak{so(3,1)^{2}}\,$ and $\,\fg=\mathfrak{so(3,1)}   \oplus_{\mathbb{Z}_{3}}  \mathfrak{u(1)^6}\,$ arising as the extensions of $\,\fg_{gauge}=\mathfrak{so(3,1)}\,$, also appear as extensions of $\,\fg_{gauge}=\mathfrak{so(4)}\,$ and $\,\fg_{gauge}=\mathfrak{iso(3)}\,$ respectively. At this point, we have to go back to section~\ref{sec:N=1}\, and emphasise that, in the T-fold description, a family of 4d effective models is determined not only by the Supergravity algebra $\,\fg\,$, but also by specifying the subalgebra $\,\fg_{gauge}\,$ associated to the isotropy subgroup of the coset space $\,\cG/\cG_{gauge}$. In this sense, effective models based on the same $\,\fg\,$, but containing different $\,\fg_{gauge}$, result in non equivalent models.


\section{The flux-induced 4d effective models}
\label{sec:pheno}

An ansatz of isotropic fluxes is compatible with vacua in which the geometric moduli, namely the complex structure and the K\"ahler moduli, are also isotropic. For the isotropic $\,\mathbb{Z}_{2} \times \mathbb{Z}_{2}\,$ orbifold, there will be one complex structure modulus and one K\"ahler modulus. Apart from those, the effective models also include the standard axiodilaton $\,S$. In this section we evaluate the flux-induced $\,\mathcal{N}=1\,$ scalar potential that determines the moduli dynamics for the set of Supergravity algebras found in the previous section. To start with, we use the T-fold description of the effective models, in which only the $\,Q\,$ and $\,\bar{H}_{3}\,$ fluxes are turned on and $\,\fg\,$ takes the form given in eqs~(\ref{IIBalgebra}).  Then, and by performing  a IIB $\leftrightarrow$ IIA mapping, these models are reinterpreted as type IIA compactifications in the presence of $\,\bar{H}_{3}\,$, $\,\omega\,$, $\,Q\,$ and $\,R\,$ fluxes.

\subsection{The canonical T-fold description}

Working in the T-fold description of the isotropic $\,\mathbb{Z}_{2} \times \mathbb{Z}_{2}\,$ orbifold models, and restricting our search to vacua with isotropic moduli
\beq
U_{1}=U_{2}=U_{3} \equiv U \hspace{1cm} \textrm{ and } \hspace{1cm} T_{1}=T_{2}=T_{3} \equiv T \ ,
\label{IsoModuli}
\eeq
there is one complex structure modulus, $U$, and one K\"ahler modulus, $T$. The $\,\bar{H}_{3}\,$ and $\,Q\,$ generalised fluxes, together with the R-R $\,\bar{F}_{3}\,$ flux, can be consistently turned on. The set $\,\left( U,S,T \right)\,$ of moduli fields obey the 4d T-duality invariant\footnote{By switching the duality frame from the T-fold to the type IIA with O6-planes description of the effective models, the K\"ahler modulus and the complex structure modulus are swapped. The models are still defined by eqs~(\ref{kwiso}). However, the coefficients in the flux-induced polynomials (\ref{P123Iso}) have to be reinterpreted in terms of the type IIA flux entries, namely, the set of $\,\bar{F}_{p}\,$ R-R fluxes with $\,p=0,2,4,6\,$ together with the entire set $\,\bar{H}_{3}\,$, $\,\omega\,$, $\,Q\,$ and $\,R\,$ of fluxes \cite{Shelton:2005cf,Aldazabal:2006up}.} Supergravity defined by the standard K\"ahler potential and the flux-induced superpotential \cite{Shelton:2005cf,Font:2008vd}
\beq
\begin{array}{ccl}
K & = & -3\,\log\left( -i\,(U-\bar{U})\right)  - \,\log\left( -i\,(S-\bar{S})\right)  - 3\,\log\left( -i\,(T-\bar{T})\right) \ , \\[3mm]
W & = &W_{RR}(U) + W_{Gen}(U,S,T) \quad .
\label{kwiso}
\end{array}
\eeq
The term $\,W_{RR}\,$ accounts for the R-R $\,\bar{F}_{3}\,$ flux-induced contribution to the superpotential and is given by
\beq
W_{RR}(U) = P_{1}(U) \ ,
\label{WRR}
\eeq
whereas the $\,\bar{H}_{3}\,$ and $\,Q\,$ fluxes induce a linear coupling for the $\,S\,$ and $\,T\,$ fields respectively, given by
\beq
W_{Gen}(U,S,T) = P_{2}(U) \, S + P_{3}(U) \, T \ .
\label{WGen}
\eeq
The flux-induced functions $\,P_{1,2,3}(U)\,$ are cubic polynomials that depend on the complex structure modulus, $U$,
\beq
\begin{array}{cclcc}
P_{1}(U) & = &  a_{0}-3\,a_{1}\,U+3\,a_{2}\,U^{2}-a_{3}\,U^{3}  &,& \\
P_{2}(U) & = & -b_{0}+3\,b_{1}\,U-3\,b_{2}\,U^{2}+b_{3}\,U^{3}  &,& \\
P_{3}(U) & = & 3\, \left(  c_{0}+ (2 c_{1}-\tilde{c}_{1}) \,U - (2c_{2}-\tilde{c}_{2})\,U^{2} - c_{3}\,U^{3} \right)   &,& 
\end{array}
\label{P123Iso}
\eeq
where the $a$s, $b$s and $c$s are coefficients that expand the $\,\bar{F}_{3}\,$, $\,\bar{H}_{3}\,$ and $\,Q\,$ fluxes respectively (see appendix).

As it was established in ref.~\cite{Font:2008vd}, performing a non linear transformation on the complex structure $\,U\,$ modulus
\beq
U \, \rightarrow \,  \mathcal{Z} \equiv \Gamma \, U = \frac{\a \, U + \b}{ \g \, U + \d} \ ,
\label{UModular}
\eeq
via the general $\,\Gamma \in \textrm{GL}(2,\mathbb{R})\,$ matrix
\beq
\Gamma \equiv \left( 
\begin{array}{cc}
 \alpha  & \beta  \\
 \gamma  & \delta
\end{array}
\right) \ ,
\label{Gammamatrix}
\eeq 
is equivalent to applying a $\,\textrm{SL}(2,\mathbb{R})\,$ rotation on the generators of the algebra (\ref{IIBalgebra}) given by
\vspace{2mm}
\beq
\left(
       \begin{array}{c}
            E^I \\
            \widetilde{E}^I
       \end{array}
\right)
= \frac{\Gamma}{|\Gamma|^{2}}\,  
\left(
       \begin{array}{c}
            - X^{2I-1} \\
              X^{2I} 
       \end{array}
\right)  \hspace{10mm} \textrm{and}  \hspace{10mm}   
\left(
       \begin{array}{c}
          -  D_I \\
             \widetilde{D}_I
       \end{array}
\right)
= \frac{\textrm{Adj}(\Gamma)}{|\Gamma|^{2}}\,  
\left(
       \begin{array}{c}
           -   Z_{2I-1} \\
                Z_{2I} 
       \end{array}
\right) \ , 
\\[2mm] 
\label{RotGen}
\eeq
for $\,I=1,2,3$. Thus, any $\,Q\,$ flux consistent with eqs~(\ref{IIBalgebra}) can always be transformed to the canonical $\,\mathcal{Q}\,$ form of eqs~(\ref{Qcanonic}), satisfying (\ref{kappaJacobi}), by means of an appropriate choice of the $\,\Gamma\,$ matrix\footnote{At this point it becomes clear that a rescaling of the gauge generators in (\ref{EGen}) is equivalent to a rescaling of the diagonal entries within the  $\,\Gamma\,$ matrix in (\ref{RotGen}). Therefore, $\,\kappa_1\,$ and $\,\kappa_2\,$ can always be expressed as their
normalised values, shown in table~\ref{tablekappa}, without lost of generality.}. The new gauge-isometry mixed brackets are still given by the co-adjoint action $\,\mathcal{Q}^{*}\,$ of $\,\mathcal{Q}\,$, and $\,\fg\,$ is forced by the $\,\mathcal{H}\mathcal{Q}=0\,$ Jacobi identity to be that of table~\ref{kappacontsol} or \ref{kappadiscsol}.  

Reading the canonical $\,\mathcal{Q}\,$ and $\,\mathcal{H}\,$ fluxes from there, and undoing the change of basis (\ref{RotGen}), we obtain the non canonical embedding of $\,\mathfrak{g}\,$ within the original $\,Q\,$ and $\,\bar{H}_{3}\,$ fluxes, respectively. Substituting them into the flux-induced polynomials within the $\,W_{Gen}\,$ piece (\ref{WGen}) of the superpotential (\ref{kwiso}), they result in
\beq
P_2(U)=(\g U + \d)^3 \cP_2(\cz) \qquad , \qquad P_3(U)=(\g U + \d)^3 \cP_3(\cz) \ ,
\label{cP23def}
\eeq 
which are functions of the transformed $\,\cZ\,$ modulus of eq.~(\ref{UModular}). The $\,\cP_2(\cz)\,$ and $\,\cP_3(\cz)\,$ flux-induced polynomials are shown in table~\ref{tablecP23}.
\begin{table}[h]
\small{
\renewcommand{\arraystretch}{1.15}
\begin{center}
\begin{tabular}{|c|c|c|}
\hline
                                      & $\cP_3(\cz)/3$ & $\cP_2(\cz)$ \\
\hline
\hline
 $\kappa_{1}=\kappa_{12}$   & $ \kappa_{2} \, \cz^3 - \kappa_{1} \, \cz$   &  $ \kappa_{2} \, \left(  \eps_{1} \, \cz^{3}  +  3  \, \eps_{2} \, \cz^{2}  \right) + \kappa_{1} \, \left( \eps_{2} +  3 \, \eps_{1} \, \cz  \right)   $   \\
\hline
 $\kappa_{12}=\kappa_{2}=0$  & $\kappa_{1} \, \cz$ & $\kappa_{1} \, \left( \eps_1 \, \cz^3 + \eps_2 \right) $ \\
\hline
\end{tabular}
\end{center}
\caption{The flux-induced polynomials.}
\label{tablecP23}
}
\end{table}

The R-R flux-induced polynomial (\ref{WRR}) in $\,W_{RR}\,$ can also be written in the form
\beq
P_1(U)=(\g U + \d)^3 \cP_1(\cz)  \ ,
\label{hatcP1def}
\eeq 
where $\,\cP_1(\cz)\,$ is a cubic polynomial that depends on $\,\cZ\,$, which can be always expanded in the basis of monomials $\,\left\lbrace 1,\cZ,\cZ^2,\cZ^3\right\rbrace$. However, it is more convenient to write  
\beq
\cP_1(\cz) = \xi_{s} \, \cP_2(\cz)  \,+\,  \xi_{t} \, \cP_3(\cz) \,-\, \xi_{3} \tilde{\cP}_{2}(\cZ) \,+\,  \xi_{7} \, \tilde{\cP}_{3}(\cZ)  \ ,
\label{cP1def}
\eeq
where $\,\tilde{\cP}_{i}(\cZ)\,$ denotes the dual of $\,\cP_{i}(\cZ)\,$ such that $\, \cP_{i} \rightarrow \frac{\tilde{\cP}_{i}}{\cZ^{3}}\,$ when $\,\cZ \rightarrow -\frac{1}{\cZ}$. This parametrization allows us to remove the R-R flux degrees of freedom, $\,\left( \xi_{s},\xi_{t}\right)\,$, from the effective theory through the real shifts
\beq
\mathcal{S} = S + \xi_{s}  \qquad , \qquad \mathcal{T}= T +  \xi_{t} \ ,
\label{shiftedfields}
\eeq
on the dilaton and the K\"ahler moduli fields \cite{Font:2008vd}. The previous argument for reabsorbing parameters fails when $\,\cP_{3}(\cz)\,$ and $\,\cP_{2}(\cz)\,$ are proportional to each other, i.e. $\,\cP_{3}(\cz) =\lambda\, \cP_{2}(\cz)\,$. In this case only the linear combination, $\,\textrm{Re}S + \lambda \, \textrm{Re}T\,$, of axions enters the superpotential, and its orthogonal direction can not be stabilised due to the form of the K\"ahler potential, see eqs~(\ref{kwiso}).

The modulus redefinition of eq.~(\ref{UModular}) translates into a transformation on the K\"ahler potential $K$ and the superpotential $W$ of the effective theory which is completely analogous to a $\,\textrm{SL}(2,\mathbb{R})_{U}\,$ modular transformation, except for a global volume factor $|\Gamma|^{3/2}$. It corresponds to a transformation $e^K|W|^2 \to e^\mathcal{K}|\mathcal{W}|^2$ of the model to an equivalent one described by the K\"ahler $\,\cK\,$ and the superpotential $\,\cW\,$ 
\beq
\begin{array}{ccl}
\mathcal{K} & = &-3 \,\log\left( -i\,(\mathcal{Z}-\bar{\mathcal{Z}})\right)  - 
\,\log\left( -i\,(\mathcal{S}-\bar{\mathcal{S}})\right)  - 3 \,\log\left(- i\,(\mathcal{T}-\bar{\mathcal{T}}) \right) \ ,  \\[2mm]
\mathcal{W}  & = &  |\Gamma|^{3/2} \left[\mathcal{T} \, \cP_3(\cz) \,+ \, \mathcal{S} \, \cP_2(\cz) -\, \xi_{3} \tilde{\cP}_{2}(\cZ) \,+\,  \xi_{7} \, \tilde{\cP}_{3}(\cZ) \right] \ ,
\label{kwModular}
\end{array}
\eeq
with the $\,\cP_{2,3}(\cZ)\,$ polynomials shown in table~\ref{tablecP23}.

One of the advantages of this parametrisation is that makes more evident the discrete symmetries of the theory.   In particular:
\begin{enumerate}
\item  $\,\cW\,$ is invariant under
\beq
\begin{array}{rclcrcl}
\cS         &  \rightarrow  & - \cS  & \hspace{5mm},& \hspace{5mm} \left( \, \eps_{1} \,,\, \eps_{2} \,,\, \xi_{3} \,,\,  \xi_{7} \, \right)      & \rightarrow  & \left( \, -\eps_{1} \,,\, -\eps_{2} \,,\, - \xi_{3} \,,\,  \xi_{7} \, \right)    \ .
\label{transWS}
\end{array}
\eeq

\item  $\,\cW\,$ goes to  $-\cW\,$ under these two transformations:
\beq
\begin{array}{rclcrcl}
 \cT        &  \rightarrow  &   -\cT &\hspace{5mm},&\hspace{5mm} \left(\, \eps_{1} \,,\, \eps_{2} \,,\, \xi_{3} \,,\, \xi_{7} \, \right)      & \rightarrow  & \left(\, -\eps_{1} \,,\, -\eps_{2} \,,\, \xi_{3} \,,\, - \xi_{7} \, \right) \ .   
\label{transWT}
\end{array}
\eeq
\beq
\begin{array}{rclcrcl}
\cZ & \rightarrow &  -  \cZ &\hspace{5mm},&\hspace{5mm}  \left(\, \eps_{1} \,,\, \eps_{2} \,,\, \xi_{3} \,,\,  \xi_{7} \, \right)    & \rightarrow &    \left(\, \eps_{1} \,,\, - \eps_{2} \,,\, - \xi_{3} \,,\,  -\xi_{7} \,\right) \ . 
\label{transWZ}
\end{array}
\eeq
\end{enumerate}
These transformations map physical vacua into non-physical ones. Using them it is possible to turn any non-physical vacuum into a physical one in a related model by flipping the signs of some parameters.  It is interesting to notice that the Supergravity algebra $\fg$ of these two models may be different (see table~\ref{tablelistalgebras}).
 
The dynamics of the moduli fields $\,\left( \cZ,\cS,\cT\right)\,$ is determined by the standard $\,\mathcal{N}=1\,$ scalar potential
\beq
V = e^\cK \left(  \sum_{\Phi=\cZ,\cS,\cT} \cK^{\Phi\bar \Phi} |D_\Phi \cW|^2 - 3|\cW|^2 \right)  \ ,
\label{VModular}
\eeq
built from eqs~(\ref{kwModular}). Since the superpotential parameters are real, the potential is invariant under field conjugation. We can combine this action with the above transformations, namely
\beq
\begin{array}{rclcrcl}
\left( \, {\cZ \,,\, \cS \,,\, \cT} \, \right) & \rightarrow &  -  \left( \, {\cZ \,,\, \cS \,,\, \cT} \, \right)^{*} &\hspace{5mm},&\hspace{5mm} \left( \, \eps_{1} \,, \, \eps_{2} \,,\, \xi_3 \,,\, \xi_7 \, \right)    & \rightarrow &    \left( \, \eps_{1} \,,\, - \eps_{2} \,,\, \xi_3 \,,\, \xi_7 \, \right) \ ,
\label{transWAll}
\end{array}
\eeq
to relate physical vacua at $\,\pm  \eps_{2}$. Notice that this transformation keeps the Supergravity algebra $\fg$ invariant.

Following the notation and conventions of ref.~\cite{Font:2008vd}, let us split the complex moduli fields into real and imaginary parts as follows 
\beq
\mathcal{Z}= x + i y \hspace{1cm},\hspace{1cm} \mathcal{S}=s + i \sigma  \hspace{1cm},\hspace{1cm}  \mathcal{T}= t+ i \mu \ .
\label{ModuliReIm}
\eeq
As in ref.~\cite{Font:2008vd}, we will adopt the conventions $\,\textrm{Im} U_{0} > 0\,$ and $\,|\Gamma| > 0\,$ without loss of generality. This implies that $\,\textrm{Im} \cZ_{0} = |\Gamma| \,\textrm{Im} U_{0} / |\gamma\,U_{0}+\delta|^{2} > 0\,$ at any physical vacuum. Moreover, the relation between the moduli VEVs and certain physical quantities like the string coupling $\,g_{s}=1/\sigma_{0}\,$ and the internal volume $\,\textrm{Vol}_{6}=\left( \mu_{0}/\sigma_{0}\right)^{3/2}\,$, imposes that $\,\sigma_{0} > 0\,$ and $\,\mu_{0} > 0 \,$ at the vacuum.

Non vanishing $\,\bar{H}_{3}\,$, $\,Q\,$ and $\,\bar{F}_{3}\,$ fluxes will generate a flux-induced tadpole for the R-R $4$-form $\,C_{4}\,$ and $8$-form $\,C_{8}\,$ potentials. Furthermore, there will be an additional $\,C_{4}\,$ tadpole due to the presence of localised O3-planes and D3-branes, as well as a $\,C_{8}\,$ tadpole due to O7-planes and D7-branes. For the scalar potential eq.~(\ref{VModular}) to contain the effect of these localised sources, both tadpoles have to be exactly cancelled. Otherwise the scalar potential could not be entirely computed from a superpotential within the $\,\mathcal{N}=1\,$ formalism. 

In the isotropic $\,\mathbb{Z}_{2} \times \mathbb{Z}_{2}\,$ orbifold, the total orientifold charge is $\,-32\,$ for the O3-planes and $\,32\,$ for the O7-planes. The flux-induced $\,C_{4}\,$ tadpole comes from the $6$-form $\,\bar{H}_{3} \wedge \bar{F}_{3} \,$ and results in the constraint 
\beq
a_{0}\,b_{3} - a_{1}\,b_{2} + a_{2}\,b_{1} - a_{3}\,b_{0}=N_{3} \ ,
\label{O3tad}
\eeq 
while that of the $\,C_{8}\,$, induced by the $2$-form $\,Q\bar{F}_{3}\,$, becomes  
\beq
a_{0}\,c_{3}+a_{1}\,\left( 2 \,c_{2}-\tilde{c}_{2} \right) - a_{2}\,\left( 2 \,c_{1}-\tilde{c}_{1} \right)-a_{3}\,c_{0} =N_{7} \ , 
\label{O7tad}
\eeq
where $N_3=32-N_{\rm D3}$ and $N_{7}=-32+N_{{\rm D7}}$. $N_{\rm D3}$ and $N_{{\rm D7}}$ are the number of D3-branes and D7-branes that are generically allowed. The original flux entries appearing in eqs~(\ref{P123Iso}) can be read from (\ref{cP23def}) and (\ref{hatcP1def}) after using (\ref{cP1def}) and substituting the $\,\cz\,$ redefined modulus of eq.~(\ref{UModular}) into the flux-induced polynomials given in table~\ref{tablecP23}. Then, the tadpole cancellation conditions result in a few simple expressions shown in table~\ref{tabletads}. As can be seen from it, the discrete transformations in 
(\ref{transWS}), (\ref{transWT}), (\ref{transWZ}) and (\ref{transWAll}) imply 
$(N_{3}, N_{7})  \rightarrow (-N_{3},N_{7})$, 
$(N_{3}, N_{7})  \rightarrow (N_{3},-N_{7})$, 
$(N_{3}, N_{7})  \rightarrow (-N_{3},-N_{7})$ and 
$(N_{3}, N_{7})  \rightarrow (N_{3},N_{7})$ respectively.
\begin{table}[htb]
\small{
\renewcommand{\arraystretch}{1.15}
\begin{center}
\begin{tabular}{|c|c|c|}
\hline
                                & $N_3/|\Gamma|^3$ & $N_7/|\Gamma|^3$ \\
\hline
\hline
$\kappa_{1}=\kappa_{12}$ & $3 \epsilon_1 \left(\kappa_1^{2}-\kappa _2^{2}\right) \xi_7 + \left( \, \epsilon _1^2  \, (3 \kappa _1^2+\kappa _2^2 )+\epsilon _2^2 \, \left(\kappa _1^2+3 \kappa _2^2\right) \, \right) \xi_3$ & $\epsilon_1 (\kappa _1^{2}-\kappa _2^{2} ) \xi_3+ ( \kappa _1^2+3 \kappa _2^2) \xi_7$ \\
\hline
$\kappa_{12}=\kappa_{2}=0$ &  $ \kappa _1^2 \, (\epsilon _1^2+\epsilon _2^2 ) \, \xi_3$  & $ \kappa _1^2 \, \xi_7$    \\
\hline
\end{tabular}
\end{center}
\caption{The R-R flux-induced tadpoles.}
\label{tabletads}
}
\end{table}

Summarizing, all the 4d effective models can be jointly described by the K\"ahler potential and the superpotential in eqs~(\ref{kwModular}) with the flux-induced polynomials presented in table~\ref{tablecP23}. They are totally defined in terms of the new $\,(\mathcal{Z},\mathcal{S},\mathcal{T})\,$ moduli fields, together with a small set of parameters
\begin{itemize}
\item[i)] ($\kappa_1 ,\, \kappa_2 \, ; \, \eps_1 ,\, \eps_2$), that determine the generalised fluxes and hence the twelve dimensional Supergravity algebra $\,\fg$. As it was previously explained, see footnote~5, $\,\kappa_1\,$ and $\,\kappa_2\,$ can always be taken to their normalised values shown in table~\ref{tablekappa} without lost of generality. 
The form of the superpotential (\ref{kwModular}) and the flux-induced polynomials in table~\ref{tablecP23} allows us to set the quantity $\,|\eps|\equiv \sqrt{\eps_{1}^2 + \eps_{2}^2}\,$ to $\,+1\,$ (provided it is non zero) by a rescaling of the $\,\cS\,$ modulus and the $\,\xi_{3}\,$ parameter. This leaves the angle defined by $\,\tan \theta_{\eps} \equiv   \displaystyle \frac{\eps_{2}}{\eps_{1}} \, $ as the only free parameter in the superpotential coming from the generalised fluxes. Using the reflection $\eps_{2} \rightarrow -\eps_{2}\,$ of (\ref{transWAll}), it is enough to evaluate it within the range $\,\theta_{\eps}\in [0,\pi]\,$ in order to cover the set of $\,\fg\,$ built from a particular $\,\fg_{gauge}$.
\item[ii)] ($\xi_3,\, \xi_7$), related to the localised O3/D3 and O7/D7 sources through the tadpole cancellation conditions presented in table~\ref{tabletads}. Similarly to the $\,\eps_{1,2}$ parameters, the (non vanishing) quantity $\,|\xi|\equiv \sqrt{ |\eps|^{2} \,\xi_{3}^{2} + \xi_{7}^{2}}\,$ can be normalised to $\,+ 1\,$, rescaling simultaneously the $\,\cS\,$ and  $\,\cT\,$ moduli fields together with the superpotential $\,\cW\,$, and keeping the angle  given by $\,|\eps|\, \tan \theta_{\xi} \equiv  \displaystyle  \frac{\xi_{7}}{\xi_{3}} \,$ as the only free parameter in the superpotential (\ref{kwModular}) coming from the R-R fluxes. 
\end{itemize}

In short, after applying modular transformations on the complex structure modulus $\,U\,$ as well as shifts and rescalings on the dilaton $\,S\,$ and the K\"ahler modulus $\,T\,$, their dynamics is totally encoded in two parameters $\,\theta_{\eps}\,$ and $\,\theta_{\xi}\,$. The former comes from the generalised fluxes while the latter comes from the R-R fluxes.

Finally, the use of this parametrization for the effective models allows us to extract an interesting result based on the following argument: it is well known that only the $\,U^{2}\,$ and $\,U^{3}\,$ couplings in the flux-induced polynomials $\,P_{2,3}(U)\,$ of eq.~(\ref{P123Iso}) come from the non-geometric $\,Q\,$ and $\,R\,$ fluxes, respectively. This is in the type IIA description of the effective models~\cite{Shelton:2005cf,Aldazabal:2006up}. 
\\[0mm]
On the other hand, provided consistent $\,Q\,$ and $\,\bar{H}_{3}\,$ fluxes in the type IIB description, their flux-induced polynomials can always be transformed to the form given in table~\ref{tablecP23} via the modular transformation $\,U \rightarrow \mathcal{Z}\,$ of eq.~(\ref{UModular}). Substituting the value of the $\,\kappa$-parameters given in table~\ref{tablekappa} into the flux-induced polynomials of table~\ref{tablecP23}, we can conclude that such quadratic and cubic couplings can be removed from the superpotential via a modular transformation for the models based on $\,\fg_{gauge}=\,\mathfrak{nil}\,$, $\,\mathfrak{iso(3)}\,$ and $\,\mathfrak{su(2)}+\mathfrak{u(1)^{3}}\,$ (if taking $\,\eps_{1}=0$). Therefore, these models can be described as geometric type IIA flux compactifications. In the   $\,\mathfrak{nil}\,$ case with $\,\kappa_{1}=\kappa_{12}=0\,$, a further $\,\cZ \rightarrow \frac{1}{-\cZ}\,$ inversion is needed in order to remove the quadratic and cubic couplings from $\,\cP_{3}(\cz)\,$ and $\,\cP_{2}(\cz)\,$. These geometric flux models do not possess supersymmetric AdS$_{4}$ vacua with all the moduli (including axionic fields) stabilised~\cite{Font:2008vd}.

\subsection{The type IIA description and no-go theorems}
\label{sec:no-go}

So far, we have been mostly centered on the T-fold description of the type II orientifolds on the isotropic $\,\mathbb{Z}_{2} \times \mathbb{Z}_{2}\,$ orbifold. This is mainly due to its suitability to classify the Supergravity algebras underlying the generalised fluxes. Any effective model in this description becomes an apparently\footnote{By apparently we mean that it may result in a geometric flux model when changing the duality frame, as we have shown in the previous section.} non-geometric model once we switch on a non vanishing $\,Q\,$ flux. By changing the duality frame it can always be mapped to a $\,\mathcal{N}=1\,$ type IIA string compactification with O6/D6 sources including the entire set of generalised fluxes, i.e. $\,\bar{H}_{3}\,$, $\,\omega\,$, $\,Q\,$ and $\,R\,$, together with a set of R-R $p$-form fluxes $\,\bar{F}_{p}\,$ with $\,p=0,2,4,6$.

In this duality frame, the dependence on the volume modulus $\,y \equiv (\textrm{Vol}_{6})^{\frac{1}{3}}\,$ and the 4d dilaton\footnote{$\,\varphi\,$ field is the ten dimensional dilaton and $\,\textrm{Vol}_{6}\,$ denotes the volume of the internal space.} $\,\tilde{\sigma} \equiv e^{-\varphi}\, \sqrt{\textrm{Vol}_{6}}\,$ of the $\,\bar{H}_{3}\,$ and $\,\bar{F}_{p}\,$ flux-induced terms in the potential, was computed in ref.~\cite{Hertzberg:2007wc} starting from the ten dimensional type IIA Supergravity action and performing dimensional reduction. Terms induced by  the $\omega$, $Q$ and $R$ fluxes were also introduced  by applying T-dualities on the $\bar{H}_{3}$ one due to the non existence of a ten dimensional Supergravity formulation of the theory in such generalised flux backgrounds.

The $\,\mathcal{N}=1\,$ scalar potential coming from these generalised type IIA flux compactifications can be split into three main contributions 
\beq
\VIIA=\VGen + \Vloc + \sum_{p=0\,(even)}^{6} V_{\bar{F}_{p}}  \ ,
\label{V's}
\eeq
with
\beq
\VGen=V_{\bar{H}_{3}} + V_{\omega} + V_{Q} + V_{R} \ .
\label{VGen}
\eeq
The latter is the potential energy induced by the set of generalised fluxes. The term $\Vloc$ accounts for the potential localised sources as O6-planes and D6-branes. $V_{\bar{H}_{3}}$ and $V_{{F}_{p}}$ account for the $\bar{H}_{3}$ and $\bar{F}_{p}$ flux-induced terms in the scalar potential. These are non negative because they come from the $|H_{3}|^{2}$ and $|F_{p}|^{2}$ terms in the ten dimensional action. $V_{\omega}$ accounts for the potential energy induced by the geometric $\,\omega\,$ flux. Finally, $V_{Q}$ and $V_{R}$ account for the contributions generated by the non-geometric $Q$ and $R$ fluxes.

Working with the $\,( y , \tilde{\sigma} )\,$ moduli fields (aka in the volume-dilaton plane limit), the power law dependence of all the terms in (\ref{V's}) on those fields, was found to be
\beq
V_{\bar{H}_{3}} \propto \tilde{\sigma}^{-2} \, y^{-3} \hspace{3mm},\hspace{3mm}V_{\omega} \propto \tilde{\sigma}^{-2} \, y^{-1} \hspace{3mm},\hspace{3mm}V_{Q} \propto \tilde{\sigma}^{-2} \, y \hspace{3mm},\hspace{3mm}V_{R} \propto  \tilde{\sigma}^{-2} \, y^{3}  \ , 
\label{V_dep_NS}
\eeq
for the set of generalised flux-induced contributions, together with
\beq
\Vloc \propto \tilde{\sigma}^{-3} \ ,
\label{V_dep_loc}
\eeq
for the potential energy induced by the localised sources and
\beq
V_{{F}_{p}} \propto \tilde{\sigma}^{-4} \, y^{3-p} \ ,
\label{V_dep_F}
\eeq
for the R-R flux-induced terms \cite{Hertzberg:2007wc}. The contributions to the scalar potential of eq.~(\ref{V's}) can be arranged as
\beq
\VIIA = A(y,M) \, \tilde{\sigma}^{-2} + B(M) \, \tilde{\sigma}^{-3} + C(y,M) \, \tilde{\sigma}^{-4} \ , 
\label{VIIAgeneric}
\eeq
where $\,M\,$ denotes the set of additional moduli fields in the model. $\,A(y,M)\,$ contains the contributions to the scalar potential resulting from the generalised fluxes. $\,B(M)\,$ accounts for the O6-planes and D6-branes contributions to the potential energy. Finally, $\,C(y,M)\,$  incorporates the terms in the scalar potential induced by the set of R-R fluxes. The explicit form of these functions depends on the features of the specific model under consideration. In contrast with previous works, our initial setup does not contain KK5-branes \cite{Silverstein:2007ac,Haque:2008jz} generating a contribution to the scalar potential that scales as the geometric $\,\omega\,$ flux-induced term of eq.~(\ref{V_dep_NS}), nor NS5-branes \cite{Hertzberg:2007wc,Haque:2008jz} that would induce a $\,\VNS5 \propto \tilde{\sigma}^{-2} y^{-2}\,$ term. We work within the framework of ref.~\cite{Caviezel:2008tf}, extended to include the set of generalised fluxes needed for restoring T-duality invariance at the 4d effective level.

\subsubsection{A simple no-go theorem in the volume-dilaton plane limit}

Using the general scaling properties of eqs~(\ref{V_dep_NS})-(\ref{V_dep_F}), it can be shown that the potential of eq.~(\ref{V's}) verifies
\beq
- \left(  y \, \frac{\partial \VIIA}{\partial y} + 3 \, \tilde{\sigma} \, \frac{\partial \VIIA}{\partial \tilde{\sigma}} \right) = 9 \, \VIIA +  \sum_{p=0\,({\rm even})}^{6} p \, V_{\bar{F}_{p}} - 2 \, V_{\omega} - 4 \, V_{Q} - 6 \, V_{R} \ .
\label{condition}
\eeq
Notice that the l.h.s of eq.~(\ref{condition}) vanishes identically at any extremum of the scalar potential yielding 
\beq
\VIIA = \dfrac{1}{9} \left( \Delta V  - \sum_{p=0\,(even)}^{6} p \, V_{\bar{F}_{p}}  \right) \ , 
\\[2mm]
\label{no-go}
\eeq
with $\,\Delta V \equiv 2 \, V_{\omega} + 4 \, V_{Q} + 6 \, V_{R}\,$. Whenever $\,V_{\bar{F}_{p}} > 0\,$ for some $\,p=0,2,4,6\,$, there can not exist dS/Mkw solutions (i.e. $\VIIA \ge 0$) unless
\beq
\Delta V \geq \sum_{p=0\,(even)}^{6} p \, V_{\bar{F}_{p}} \ . 
\label{muchrest}
\eeq
This was the line followed in \cite{Silverstein:2007ac,Haque:2008jz} where certain type IIA flux compactifications on curved internal spaces generating a contribution $\,\Delta V = 2 \, V_{\omega} \neq 0\,$ were presented\footnote{Building the linear combination 
\beq
- \left(  y \, \frac{\partial \VIIA}{\partial y} + \, \tilde{\sigma} \, \frac{\partial \VIIA}{\partial \tilde{\sigma}} \right) = 3 \, \VIIA +  2 \, \left(  V_{\bar{H}_{3}} + V_{\bar{F}_{4}} + 2 \, V_{\bar{F}_{6}} \right) -2 \left( V_{\bar{F}_{0}} + V_{Q} + 2\, V_{R} \right) \ ,
\label{Romancondition}
\eeq
shows that $\, V_{\bar{F}_{0}} \neq 0\,$ for dS vacua to exist in any geometric model, i.e. $\,V_{Q}=V_{R}=0$. This implies having a Romans massive Supergravity, as it was stated in \cite{Haque:2008jz}.}.

\subsection{From the T-fold to the type IIA description}
\label{sec:fromIIBtoIIA}

T-duality shuffles the generalised flux entries in going from one duality frame to other, according to eq.~(\ref{Tdualitychain}). The same happens for the R-R fluxes; the $\,\bar{F}_{3}\,$ flux entries in the T-fold description go to the different $\,\bar{F}_{p}\,$ flux entries in the type IIA one \cite{Shelton:2005cf,Aldazabal:2006up}. Since the flux-induced terms in the scalar potential map between duality frames, there should also be a mapping of the remaining contributions, namely, those coming from localised sources in both descriptions.

In the T-fold description, the scalar potential eq.~(\ref{VModular}) can be entirely computed from eqs~(\ref{kwModular}) using the flux-induced polynomials in table~\ref{tablecP23}. The contributions coming from the O3/D3 and O7/D7 localised sources are given by
\beq
\begin{array}{ccc}
\3loc &=&  - \dfrac{N_3}{16 \, \mu^{3}} \ , \\[4mm]
\7loc &=&  \dfrac{3 \, N_7}{16\,\mu^{2} \,\sigma} \ ,
\label{V37}
\end{array}
\eeq
where we have made use of the tadpole cancellation conditions shown in table~\ref{tabletads}. The $\,\mathcal{N}=1\,$ scalar potential computed from (\ref{VModular}) contains the localised sources needed to cancel the flux-induced tadpoles \cite{Villadoro:2005cu}. We will not consider additional localised sources whose effect would have to be included directly in the scalar potential \cite{Kachru:2003aw}. Therefore, all the contributions to $\,\VT\,$ (the scalar potential computed in type IIB with O3/O7-planes) coming from localised sources are those of eqs~(\ref{V37}).

By applying the relation between the IIB/IIA moduli fields for the Supergravity models based on the $\,\mathbb{Z}_{2} \times \mathbb{Z}_{2}\,$ isotropic orbifold \cite{Villadoro:2005cu},
\beq
\begin{array}{ccc}
\textrm{T-fold description}  &  \leftrightarrow &   \textrm{Type IIA description} \\[1mm]
        \sigma = \dfrac{\tilde{\sigma}}{\tilde{\mu}^{3}}        &  \leftrightarrow &  \, \tilde{\sigma}  \\[4mm]
        \mu = \tilde{\sigma} \, \tilde{\mu}         &  \leftrightarrow &   \tilde{\mu}  
\end{array}
\label{muredef}
\eeq
the $\,\3loc\,$ and $\,\7loc\,$ contributions of eqs~(\ref{V37}) turn out to depend on the $\,\tilde{\sigma}\,$ modulus in the same way as $\,\Vloc\,$ in eq.~(\ref{V_dep_loc}).

Computing the IIB with O3/O7-planes scalar potential from eqs~(\ref{kwModular}) with the flux-induced polynomials given in table~\ref{tablecP23}, and using the moduli relation of eq.~(\ref{muredef}), we end up with the standard form (\ref{VIIAgeneric}) of the scalar potential in the IIA with O6-planes language,
\beq
\VIIA = A(y,\tilde{\mu},x) \, \tilde{\sigma}^{-2} + B(\tilde{\mu}) \, \tilde{\sigma}^{-3} + C(y,\phi) \, \tilde{\sigma}^{-4} \ ,
\label{potentialIIA}
\eeq
where $\phi$ denotes the whole set of axions ($x,s,t$). The functions $A$, $B$ and $C$ account for the sixteen different sources of potential energy which we list below

\begin{itemize}

\item $A(y,\tilde{\mu},x)$ contains the contributions coming from the set of $\,R\,$, $\,Q\,$, $\,\omega\,$ and $\,\bar{H}_{3}\,$ fluxes,
\beq
\begin{array}{lcl}
A(y,\tilde{\mu},x) &=& y^{3} \left( \dfrac{r_{1}^{2}}{\tilde{\mu}^{6}} + r_{2}^{2} \, \tilde{\mu}^{2} \right) + y \left(\dfrac{q_{1}^{2}}{\tilde{\mu}^{6}} + \dfrac{q_{2}}{\tilde{\mu}^{2}} + q_{3} \, \tilde{\mu}^{2} \right) + \\[4mm]
&+& \dfrac{1}{y} \left( \dfrac{\omega_{1}^{2}}{\tilde{\mu}^{6}} + \dfrac{\omega_{2}}{\tilde{\mu}^{2}} + \omega_{3} \, \tilde{\mu}^{2} \right) + \dfrac{1}{y^{3}} \left( \dfrac{h_{1}^{2}}{\tilde{\mu}^{6}} +  h_{2}^{2} \, \tilde{\mu}^{2} \right)\ . 
\end{array}
\label{fuctiona}
\eeq

\item $B(\tilde{\mu})$ accounts for the potential energy stored within the O6-planes and D6-branes localised sources,
\beq
B(\tilde{\mu}) =  \frac{-1}{16} \left( \frac{N_{3}}{\tilde{\mu}^{3}}  -  3 \, N_{7} \, \tilde{\mu} \right)  \ .
\label{fuctionb}
\eeq 
The O3/D3 sources in the T-fold description can be interpreted in the type IIA language as O6/D6 sources wrapping a three cycle of the internal space, which is invariant under the IIA orientifold action (\ref{osigmaA})\footnote{In the type IIA language, only the O6/D6 sources wrapping this invariant three cycle preserve $\,\mathcal{N}=4\,$ supersymmetry \cite{Dall'Agata:2009gv}. Since these sources are reinterpreted as O3/D3 sources in the type IIB language, the Jacobi identities descending from a truncation of a $\,\mathcal{N}=4\,$ Supergravity algebra would have nothing to say about their number \cite{Aldazabal:2008zza}.}. However, the O7/D7 sources in the T-fold description have to be understood in the type IIA picture as O6/D6 sources wrapping three cycles which are invariant under the composition of both the IIA orientifold together with the $\,\mathbb{Z}_{2} \times \mathbb{Z}_{2}\,$ orbifold actions \cite{Aldazabal:2006up}.

\item $C(y,\phi)$ contains the terms in the scalar potential induced by the $\,\bar{F}_{p}\,$ R-R $p$-form fluxes with $\,p=0,2,4$ and $6\,$,
\beq
C(y,\phi) = y^{3} \, f_{0}^{2} + y \, f_{2}^{2} + \frac{f_{4}^{2}}{y}  + \frac{f_{6}^{2}}{y^{3}} \ .
\label{fuctionc}
\eeq 
\end{itemize}

Taking a look at the $(\tilde{\sigma},y)$-scaling properties of the different terms appearing in these functions, they are easily identified in the type IIA picture of eqs~(\ref{V_dep_NS})-(\ref{V_dep_F}), resulting in a dictionary between both descriptions at the level of the scalar potential. In fact,
\beq
\VT \leftrightarrow \VIIA  \ ,
\label{matching}
\eeq
after applying (\ref{muredef}) and reinterpreting the different scalar potential contributions in the T-fold description with respect to the type IIA duality frame. 

All the terms in the scalar potential $\,\VIIA\,$ of (\ref{V's}) and (\ref{VGen}) are reproduced. Making their dependence on the axions explicit, they are given by 
\beq
\begin{array}{lcr}
V_{\omega}  = \tilde{\sigma}^{-2}  y^{-1}  \tilde{\mu}^{-6} \Big( \omega_{1}^{2}(x) + \omega_{2}(x)  \, \tilde{\mu}^{4}+ \omega_{3}(x) \, \tilde{\mu}^{8} \Big) &,& V_{\bar{H}_{3}} =  \tilde{\sigma}^{-2}  y^{-3}  \tilde{\mu}^{-6} \Big( h_{1}^{2}(x) + h_{2}^{2}(x)  \,  \tilde{\mu}^{8} \Big) \ ,\\[8mm]
V_{Q} = \tilde{\sigma}^{-2}  y  \, \tilde{\mu}^{-6} \Big( q_{1}^{2}(x) + q_{2}(x) \, \tilde{\mu}^{4} + q_{3}(x)  \, \tilde{\mu}^{8} \Big)&,&V_{R} = \tilde{\sigma}^{-2}  y^{3}  \, \tilde{\mu}^{-6} \Big( r_{1}^{2}(x) + r_{2}^{2}(x) \, \tilde{\mu}^{8} \Big) \ ,
\label{V_Gen_NS}
\end{array}
\eeq
for the set of generalised flux-induced terms,
\beq
\Vloc = - \frac{1}{16} \, \tilde{\sigma}^{-3} \, \tilde{\mu}^{-3} \, \left( N_{3}  -  3 \, N_{7} \, \tilde{\mu}^{4} \right) \ ,
\label{V_Gen_loc}
\eeq
for the potential energy within the O6/D6 localised sources and
\beq
V_{{F}_{p}} = \tilde{\sigma}^{-4} \, y^{(3-p)} \, f_{p}^{2}(x,s,t) \hspace{10mm} , \hspace{10mm} p=0,2,4,6 \ , 
\label{V_Gen_F}
\eeq
for the R-R flux-induced contributions. The above decomposition of the scalar potential holds after the non linear action of $\,\Theta \in \textrm{GL}(2,\mathbb{R})_{\cZ}\,$ on the redefined complex structure modulus $\,\cZ \rightarrow \Theta^{-1} \, \cZ$. 

$V_{\bar{H}_{3}}$, $V_{\omega}$, $V_{Q}$ and $V_{R}$  involve just the axion $\,x = \textrm{Re}\cZ\,$ unlike the set of $\,V_{\bar{F}_{p}}\,$ contributions that depend on the entire set of them. Specifically, the functions $\,f_{p}(x,s,t)\,$ have a linear dependence on the axions $\,s\,$ and $\,t$. It is clear from (\ref{V_Gen_NS}) and (\ref{V_Gen_F}) that $\,V_{\bar{H}_{3}},V_{R},V_{\bar{F}_{p}} \,$ are positive definite, as well as the $\,V_{\omega_{1}}\,$ and $\,V_{q_{1}}\,$ terms induced by $\,\omega_{1}\,$ and $\,q_{1}\,$ respectively. 

At this point we would like to make a rough comparison of the scalar potential (\ref{potentialIIA}), involving the entire set of moduli fields, with that of ref.~\cite{Haque:2008jz} obtained in the volume-dilaton two (non-axionic) moduli limit. First of all, the compactifications studied there do not include non-geometric fluxes, i.e. $V_{Q}=V_{R}=0$, so $\,V_{\omega} \ge 0\,$ at any dS/Mkw vacuum. The setup in ref.~\cite{Haque:2008jz} also reduces the contributions in (\ref{V_Gen_loc}), accounting for localised sources, to the piece involving $\,N_{3}\,$ (with $N_{3}>0$). Finally, another difference is that the functions $\,r_{1,2}\,,\,q_{1,2,3}\,,\,\omega_{1,2,3}\,,\,h_{1,2}\,$ and $\,f_{p}\,$ in (\ref{V_Gen_NS}) and (\ref{V_Gen_F}) can not be taken to be constant \cite{Haque:2008jz}, but they do depend on the set of axions, $\,\phi$. Hence these are dynamical quantities to be determined by the moduli VEVs.

\section{Where to look for dS/Mkw vacua?}

Armed with the mapping between the T-fold and the type IIA descriptions of the effective models presented in the previous section, we investigate now how the no-go theorem of eq.~(\ref{muchrest}), on the existence of dS/Mkw vacua, can be used in this context. We restrict ourselves to vacua with {\em all}  moduli (including axions) stabilised by fluxes. We do not consider the limiting  cases defined by
\begin{enumerate}
 \item $\,\eps_{1}=\eps_{2}=0$, for which $\,\cP_{2}(\cZ)=0\,$ and the shifted dilaton $\,\cS\,$ can not be stabilised by the fluxes. This {\bf excludes algebras $\,2\,$, $\,4\,$ and $\,17\,$} in table~\ref{tablelistalgebras}. 
 \item $\,\kappa_{1}=\kappa_{2}=0$, yielding $\,\cP_{3}(\cZ)=0\,$ and leaving the shifted K\"ahler modulus $\,\cT\,$ not stabilised. This case results in no-scale Supergravity models previously found \cite{Derendinger:2004jn,Camara:2005dc}, since the $\,\cT\,$ modulus does not enter the superpotential in eq.~(\ref{kwModular}). This {\bf discards algebra $\,20\,$} in table~\ref{tablelistalgebras}.
\end{enumerate}
Furthermore, we will also assume that $\,\bar{F}_{p} \neq 0\,$ for some $\,p=0,2,4,6\,$. However, we will consider a much weaker version of the no-go theorem of eq.~(\ref{muchrest}), given by
\beq
\Delta V  > 0 \ . 
\label{lessno-go}
\eeq
The reason for doing this is that our classification of the Supergravity algebras, which is the building block for finding vacua, has nothing to do with R-R fluxes\footnote{In this work we have used the IIB with O3/O7 Supergravity algebra given in eq.~(\ref{IIBalgebra}) and proposed in ref.~\cite{Shelton:2005cf}. It is totally specified by the non-geometric $\,Q \,$ and the $\,\bar{H}_{3}\,$ fluxes. In the most recent article of ref.~\cite{Aldazabal:2008zza}, the origin of these IIB generalised flux models as gaugings of $\,\mathcal{N}=4\,$ Supergravity was explored. The R-R $\,\bar{F}_{3}\,$ flux was found to also enter the $\,\mathcal{N}=4\,$ Supergravity algebra, written this time in terms of both electric and magnetic gauge/isometry generators.}. These will not be used in the process of excluding algebras through the no-go theorem, and, therefore, we will use eq.~(\ref{lessno-go}) (rather than (\ref{muchrest})) in what follows. Note that, in any case, the R-R fluxes defining the $\,V_{\bar{F}_{p}}\,$ contributions in (\ref{V_Gen_F}) will play a crucial role in the stabilisation of the axions~\cite{deCarlos:2009qm}.

Working with the flux-induced polynomials $\,\cP_{2}(\mathcal{Z})\,$ and $\,\cP_{3}(\mathcal{Z})\,$, from table~\ref{tablecP23}, corresponds to defining the Supergravity algebra $\,\fg\,$ in the canonical basis of tables~\ref{kappacontsol} and \ref{kappadiscsol}. In this basis, $\,\Delta V\,$ reads
\beq
\Delta V = \frac{3 \, |\Gamma|^{3}}{16  \, y \, \tilde{\sigma}^{2} \, \tilde{\mu}^{6}} \, \left(   l_{2} \, \tilde{\mu}^{8}+  l_{1} \, \tilde{\mu}^{4} + l_{0} \right) \hspace{1cm} \textrm{ where } \hspace{1cm} |\Gamma| \, , \, l_{0} >0 \ .
\\[2mm]
\label{DeltaVPoly}
\eeq
The functions $\,l_{2}\,$, $\,l_{1}\,$ and $\,l_{0}\,$ in the polynomial of (\ref{DeltaVPoly}) may depend on the $\,\cZ\,$ modulus and determine whether or not $\,\Delta V\,$ can be positive (provided that $\,y_{0},\tilde{\sigma}_{0},\tilde{\mu}_{0} > 0\,$ at any physical vacuum). 

In some cases, moving to a different algebra basis may simplify the flux-induced polynomials in the superpotential, since they are built from the structure constants of $\,\fg$. Then, a higher number of zero entries in the structure constants translates into simpler effective models. This also simplifies the $\,l_{2}\,$, $\,l_{1}\,$ and $\,l_{0}\,$ functions in eq.~(\ref{DeltaVPoly}), which determine whether the necessary condition (\ref{lessno-go}) can be fulfilled.

Starting with the effective theory derived in the canonical basis of $\,\fg\,$, and by applying a non linear $\, \Theta \in \textrm{GL}(2,\mathbb{R})_{\cZ}\,$ transformation upon the $\,\cZ\,$ modulus, $\,\cZ \rightarrow \Theta^{-1} \, \cZ \,$, we end up with an equivalent effective theory formulated in a non canonical basis. The generators in this new basis are related to the original $\,X^{a}\,$ and $\,Z_{a}\,$ through the same rotation of (\ref{RotGen}), by simply replacing
\beq
\Gamma \rightarrow \Theta\,\Gamma \ .
\label{Gamma'}
\eeq
Since the scalar potential decomposition introduced in section~\ref{sec:fromIIBtoIIA} holds after the $\,\cZ \rightarrow \Theta^{-1} \, \cZ\,$ transformation, the form of the $\,\Delta V\,$ in (\ref{DeltaVPoly}) also does. Therefore, restricting the $\,\Theta\,$ transformations to those with $\,|\Theta|>0$, i.e. $\textrm{Im}\cZ_{0}>0 \rightarrow \textrm{Im}(\Theta^{-1} \, \cZ_{0})>0$, guarantees that (\ref{lessno-go}) still holds as a necessary condition for dS/Mkw vacua to exist.

The usefulness of moving from the canonical basis to a non-canonical one can be illustrated in the two particular cases determined by the transformations
\vspace{2mm}
\beq
\Theta_{1} \equiv \left( 
\begin{array}{cr}
 0  & -1  \\
 1  &  0
\end{array}
\right)
\hspace{8mm} \textrm{and} \hspace{8mm}
\Theta_{2} \equiv \frac{1}{2^{2/3}} \left( 
\begin{array}{cr}
 1  & -1  \\
 1  & 1
\end{array}
\right) \ .
\\[2mm]
\label{Thetamatrices}
\eeq
$\Theta_{1}\,$ exchanges the gauge generators $\,E^{I} \leftrightarrow \wt E^{I}\,$ as well as the isometry ones $\,D_{I} \leftrightarrow \wt D_{I}\,$ (up to signs). On the other hand, $\,\Theta_{2}\,$ induces the well known rotation needed for turning the $\,\mathfrak{so(4)}\,$ algebra into the direct sum of $\,\mathfrak{su(2)^{2}}\,$ in both the gauge and isometry subspaces.\\

\vspace{6mm}
\textbf{Using the canonical basis.}\\

Let us start by exploring the existence of dS/Mkw vacua in two sets of effective models computed in the canonical basis of $\,\fg$:
\begin{enumerate}
\item Taking the $\,\kappa_{1}=\kappa_{12}\,$ solution of eq.~(\ref{kappaJacobi}) and  fixing $\,\kappa_{2}=0\,$,  results in $\,V_{Q}=V_{R}=0\,$. These models are based on $\,\fg_{gauge}=\mathfrak{iso(3)}\,$ and admit a geometric type IIA description. The coefficients determining the quadratic polynomial in (\ref{DeltaVPoly}) are given by
\beq
l_{2} = - \kappa_{1}^{2} \hspace{5mm} , \hspace{5mm}     l_{1} = 4 \, \eps_{1} \, \kappa_{1}^{2}         \hspace{5mm} \textrm{and} \hspace{5mm} l_{0}= \eps_{1}^{2} \, \kappa_{1}^{2} \ .
\eeq
The case with $\,\eps_{1}=0\,$ translates into $\,\Delta V < 0$, so dS/Mkw solutions are {\bf forbidden for  algebra $\,10\,$} in table~\ref{tablelistalgebras}. This Supergravity algebra has received special attention in ref.~\cite{Dall'Agata:2009gv}, where it has been identified as $\,\fg=\mathfrak{su(2) \otimes_{\mathbb{Z}_{3}}\mathfrak{n_{9,3}}}\,$. In fact, fixing $\,\kappa_{2}=\eps_{1}=0\,$ in the commutation relations of table~\ref{kappacontsol}, the algebra is given by
\beq
\begin{array}{lcc}
\left[ E^{I},E^{J}\right]=\eps_{IJK} E^{K} & , & \left[ E^{I},A_{n}^{J} \right]=\eps_{IJK} A_{n}^{K} \;, \\
\left[ A_{1}^{I},A_{1}^{J} \right]=\eps_{2}\,\, \eps_{IJK} A_{2}^{K} & , & \left[ A_{1}^{I},A_{2}^{J} \right]=\eps_{IJK} A_{3}^{K} \;,
\label{algebraexample}
\end{array}
\eeq
with $n=1,2,3$, after identifying $\,A_{1}\equiv \wt D$, $A_{2}\equiv -\wt E\,$ and $\,A_{3}\equiv D$. It coincides with that of \cite{Dall'Agata:2009gv}\footnote{Observe that the $\,(A^{I}_{2},A^{I}_{3})_{I=1,2,3}\,$ generators expand a $\,\mathfrak{u}(1)^{6}\,$ abelian ideal in the algebra (\ref{algebraexample}). After taking the quotient by this abelian ideal, the resulting algebra involving the $\,(E^{I},A^{I}_{1})_{I=1,2,3}\,$ generators becomes $\,\mathfrak{iso(3)}$, so (\ref{algebraexample}) is equivalent to $\,\mathfrak{g}=\mathfrak{iso(3)} \oplus_{\mathbb{Z}_{3}} \mathfrak{u(1)^{6}}\,$ as it was identified in table~\ref{tablelistalgebras}.}, and we can now exclude that it has any  dS/Mkw vacua\footnote{We are always under the assumption of isotropy on both flux backgrounds and moduli VEVs.}. Moreover, if $\,\eps_{1} \neq 0\,$, the case $\,\eps_{2}=0\,$ cannot have all the axions stabilised, since only the linear combination $\,\textrm{Re}S - \, \eps_{1}^{-1} \, \textrm{Re}T\,$ enters the superpotential.

\item Taking  $\,\kappa_{12}=\kappa_{2}=0\,$  in eq.~(\ref{kappaJacobi}) induces non-geometric $\,V_{Q}\neq 0\,$ and $\,V_{R}\neq 0\,$ contributions in the scalar potential. These models are built from $\,\fg_{gauge}=\mathfrak{su(2)}+\mathfrak{u(1)^{3}}\,$ and the quadratic polynomial in (\ref{DeltaVPoly}) results in
\beq
l_{2} = - \kappa_{1}^{2} \hspace{5mm} , \hspace{5mm}     l_{1} = 2 \, \eps_{1} \, \kappa_{1}^{2}\, \left(|\cZ|^{2} + (\textrm{Im}\cZ)^{2} \right)         \hspace{5mm} \textrm{and} \hspace{5mm} l_{0}= \eps_{1}^{2} \, \kappa_{1}^{2} \, |\cZ|^{4}                \ .
\eeq
As in the previous case, the limit $\,\eps_{1}=0\,$ yields effective models with $\,\Delta V < 0\,$ as well as $\,V_{Q}=V_{R}=0\,$. They also admit to be described as geometric type IIA flux compactifications where dS/Mkw solutions are again forbidden. Hence the {\bf exclusion of  algebras $\,16\,$ and $\,17\,$} in table~\ref{tablelistalgebras}.
\end{enumerate}

\vspace{6mm}
\textbf{Using the $\,\Theta_{1}$-transformed basis.}\\

Leaving the canonical basis via applying the $\,\Theta_{1}\,$ transformation in eq.~(\ref{Thetamatrices}), additional effective models can be excluded from having vacua with non-negative energy:

\begin{enumerate}
\item Taking $\,\kappa_{1}=\kappa_{12}$ in eq.~(\ref{kappaJacobi}), specifically $\,\kappa_{1}=\kappa_{12}=0$, the effective models are those based on $\,\fg_{gauge}=\mathfrak{nil}\,$. Condition (\ref{lessno-go}) is not efficient in excluding the existence of dS/Mkw vacua in any region of the parameter space when working in the canonical basis of $\,\fg\,$.
 
Applying the $\,\Theta_{1}^{-1}\,$ transformation of $\,\cZ \rightarrow \frac{1}{-\cZ}$, the flux-induced polynomials get simplified to
\beq
\cP_{3}(\cZ)= 3 \, \kappa_{2}   \hspace{5mm},\hspace{5mm} \cP_{2}(\cZ)=\kappa_{2} \, \left(  \eps_{1} - 3\, \eps_{2} \, \cz   \right) \ ,
\eeq
having lower degree than their canonical version shown in table~\ref{tablecP23}. In this new basis, the non-geometric contributions to the scalar potential identically vanish, $\,V_{Q}=V_{R}=0\,$, so these effective models can eventually be described as geometric type IIA flux compactifications. The coefficients determining the quadratic polynomial in (\ref{DeltaVPoly}) are now given by
\beq
l_{2} = 0 \hspace{5mm} , \hspace{5mm}     l_{1} = 0       \hspace{5mm} \textrm{and} \hspace{5mm} l_{0} = \eps_{2}^{2} \, \kappa_{2}^{2} \ .
\eeq
Therefore,  condition (\ref{lessno-go}) excludes the existence of dS/Mkw vacua in the limit case of $\,\eps_{2}=0\,$ since $\,\Delta V = 0\,$. This is {\bf algebra $\,13\,$} in table~\ref{tablelistalgebras}.

\item Taking  $\,\kappa_{12}=\kappa_{2}=0\,$ in eq.~(\ref{kappaJacobi}),  $\,\fg_{gauge}=\mathfrak{su(2)}+\mathfrak{u(1)^{3}}\,$. The resulting effective models were previously explored in the canonical basis of $\,\fg\,$, discarding the existence of dS/Mkw solutions if $\,\eps_{1}=0\,$. Applying again the $\,\Theta_{1}^{-1}\,$ inversion of $\,\cZ \rightarrow \frac{1}{-\cZ}\,$, the flux-induced polynomials result in
\beq
\cP_{3}(\cZ)= 3 \, \kappa_{1} \, \cZ^{2}   \hspace{5mm},\hspace{5mm} \cP_{2}(\cZ)=\kappa_{1} \, \left(  \eps_{1} -  \eps_{2} \, \cz^{3}   \right) \ ,
\eeq
and the coefficients in the quadratic polynomial in (\ref{DeltaVPoly}) are modified to
\beq
l_{2} = - 2\, \kappa_{1}^{2} \, (\textrm{Im}\cZ)^{2} \hspace{5mm} , \hspace{5mm}     l_{1} = 2 \, \kappa_{1}^{2} \, \eps_{1}  \hspace{5mm} \textrm{and} \hspace{5mm} l_{0} = \eps_{2}^{2} \, \kappa_{1}^{2} \, |\cZ |^{4} \ .
\eeq
dS/Mkw vacua are automatically forbidden for $\,\eps_{2} = 0\,$ as long as $\,\eps_{1} \leq 0\,$, corresponding to {\bf algebras $\,17\,$ and $\,19\,$} in table~\ref{tablelistalgebras}.
\end{enumerate}

\vspace{6mm}
\textbf{Using the $\,\Theta_{2}$-transformed basis.}\\

The last family of effective models to which  eq.~(\ref{lessno-go}) applies in a useful way is that coming from fixing $\,\kappa_{1}=\kappa_{12}\,$, specifically $\,\kappa_{1}=\kappa_{12}=\kappa_{2}=\kappa\,$ in (\ref{kappaJacobi}). It implies $\,\fg_{gauge}=\mathfrak{so(4)}\,$. Performing this time the $\,\Theta_{2}^{-1}\,$ transformation of $\,\cZ \rightarrow \frac{1}{2^{1/3}} \left( \frac{\,\,\,\,\,\cZ + 1}{-\cZ + 1} \right)$, the flux-induced polynomials simplify to
\beq
\cP_{3}(\cZ)= 3 \, \kappa \, \cZ \, \left(1 + \cz \right)  \hspace{5mm},\hspace{5mm} \cP_{2}(\cZ)= \kappa \, \left( \eps_{-} \, \cZ^{3}  + \eps_{+} \right) \ ,
\label{cP23Shift}
\eeq
where $\,\eps_{\pm}=\eps_{1}\pm\eps_{2}\,$. This $\,\Theta_{2}$-induced transformation splits $\,\fg\,$ into the direct sum of two six dimensional pieces, $\,\fg=\fg_{+}+\fg_{-}\,$, determined by the sign of the $\,\eps_{\pm}\,$ parameters, respectively. $\,\fg\,$ acquires a manifest $\,\mathbb{Z}_{2} \oplus \mathbb{Z}_{2}$ graded structure. These effective models result with a symmetry under the exchange of $\,\eps_{-} $ and $\eps_{+}\,$ \cite{Font:2008vd}. In fact, the effective action becomes invariant under this swap, together with the moduli redefinitions 
\beq
\cZ \rightarrow 1/\cZ^{*} \hspace{4mm},\hspace{4mm} \cS \rightarrow - \cS^{*} \hspace{4mm},\hspace{4mm} \cT \rightarrow -\cT^{*} \ .
\label{extrasym}
\eeq 
This symmetry comes from combining eq.~(\ref{transWAll}) with $\Theta_{2}^{-1}$.
\\[1mm]
Working out the scalar potential, using (\ref{cP23Shift}), the coefficients in the quadratic polynomial (\ref{DeltaVPoly}) take the form 
\beq
l_{2} = - \kappa^{2} \, \left( 1 + 2  \, (\textrm{Im}\cZ)^{2} \right) \hspace{2mm} , \hspace{2mm}     l_{1} = 2 \, \kappa^{2} \, \left( \eps_{-} \, \left( |\cZ|^{2} + (\textrm{Im}\cZ)^{2} \right) + \eps_{+} \right)       \hspace{3mm} \textrm{and} \hspace{3mm} l_{0} = \eps_{-}^{2} \, \kappa^{2} \, |\cZ |^{4} \ ,
\label{l2l1l0so4}
\eeq
and physically viable dS/Mkw vacua are excluded in the limiting case $\,\eps_{-}=0$ as long as $\,\eps_{+} \leq 0$. 
The invariance of the effective action under the exchange of $\,\eps_{-} $ and $\eps_{+}\,$ together with the moduli redefinitions of (\ref{extrasym}), implies that  ($\,\eps_{+}=0\,$, $\,\eps_{-} \leq 0$) is also excluded. These are {\bf algebras $\,4\,$ and $\,8\,$} in table~\ref{tablelistalgebras}.\\

\vspace{6mm}
\textbf{Collecting the results.}\\

Finally, the effective models with $\,\kappa_{1}=\kappa_{12}=-\kappa_{2}=\kappa\,$ built from $\,\fg_{gauge}=\mathfrak{so(3,1)}\,$ can not be ruled out and may have dS/Mkw vacua at any point in the parameter space.
Three main results can be highlighted for our isotropic orbifold, also assuming isotropic VEVs for the moduli:
\begin{itemize}
\item Eight of the twenty algebra-based effective models admit a geometrical description as a type IIA flux compactifications, whereas the remaining twelve are forced to be non-geometric flux compactifications in any duality frame. 

\item The four effective models based on the semisimple Supergravity algebras $\,1,3,5$ and $7\,$ are non-geometric flux compactifications in any duality frame.

\item No effective model based on a semisimple $\,\fg\,$ satisfying (\ref{semisimplecond}), can be excluded from having dS/Mkw vacua using (\ref{lessno-go}). On the other hand, more than half of the effective models based on non-semisimple Supergravity algebras can be discarded. 
\end{itemize}

These results are presented in table~\ref{tableexcluded} which complements the previous table~\ref{tablelistalgebras} in characterising the set of non equivalent effective models. 
\begin{table}[htb]
\small{
\renewcommand{\arraystretch}{1.15}
\begin{center}
\begin{tabular}{|c|p{3.37mm}|p{3.37mm}|p{3.37mm}|p{3.37mm}|p{3.37mm}|p{3.37mm}|p{3.37mm}|p{3.37mm}|p{3.37mm}|p{3.37mm}|p{3.37mm}|p{3.37mm}|p{3.37mm}|p{3.37mm}|p{3.37mm}|p{3.37mm}|p{3.37mm}|p{3.37mm}|p{3.37mm}|p{3.37mm}|}
\hline
$\#$ & $\,1$ & $\,2$ & $\,3$ & $\,4$ & $\,5$ & $\,6$ & $\,7$ & $\,8$ & $\,9$ & $10$ & $11$ & $12$ & $13$ & $14$ & $15$ & $16$ & $17$ & $18$ & $19$ & $20$ \\
\hline
\hline
\textrm{\begin{footnotesize}\textsc{class}\end{footnotesize}}   & $\! \fnNG$ & $\! \fnNG$ & $\! \fnNG$ & $\! \fnNG$ & $\! \fnNG$ & $\! \fnNG$ & $\! \fnNG$ & $\! \fnNG$ & $\,\fnG$ & $\,\fnG$ & $\,\fnG$ & $\,\fnG$ & $\,\fnG$ & $\! \fnNG$ & $\! \fnNG$ & $\,\fnG$ & $\,\fnG$ & $\! \fnNG$ & $\! \fnNG$ & $\,\fnG$ \\
\hline
\hline
$V_{0} \ge 0$   & $\checkmark$ & $\:\times$ & $\checkmark$ & $\:\times$ & $\checkmark$ & $\checkmark$ & $\checkmark$ & $\:\times$ & $\checkmark$ & $\:\times$ & $\checkmark$ & $\checkmark$ & $\:\times$ & $\checkmark$ & $\checkmark$ & $\:\times$ & $\:\times$ & $\checkmark$ & $\:\times$ & $\:\times$ \\
\hline
\end{tabular}
\end{center}
\caption{Splitting of the algebra-based effective models into two classes: those admitting to be described as geometric (G) flux backgrounds by changing the duality frame, and those being non-geometric (NG) flux backgrounds in any duality frame. The mark $\,\times\,$ indicates that the model is excluded by the necessary condition ($4.1$) from having dS/Mkw vacua ($V_{0} \ge 0$),  whereas if it is not,  we use the label $\,\checkmark$.}  
\label{tableexcluded}
}
\end{table}
\\

Let us analyse a concrete example to illustrate the usefulness of this table. We consider an effective model in terms of the original $(U,S,T)$ moduli fields with the standard K\"ahler potential of eq.~(\ref{kwiso}), and with superpotential
\beq
\begin{array}{ccl}
W(U,S,T) &=& 6 \, T \, \left(U^3+U^2-U-1\right) + 2 \, S \, \left(U^3+3 \, U^2+3 \, U+ 1 \right) \, + \\ 
& + & 2 \, \left(-U^3+3 \, U^2-3 \, U-3 \right) \ ,
\end{array}
\label{Wexample}
\eeq
where the flux entries are given by $\,\tilde{c}_1=\tilde{c}_2=c_{i} =-2 \,$, $i=0,..,3$ for the non-geometric $Q$ flux and
$b_{0}=b_{2}=-b_{1}=-b_{3} = -2$,   $a_{0}=-3 a_{1}=-3 a_{2}=-3 a_{3}=-6$ for the NS-NS and the R-R fluxes. This set of flux coefficients are even and satisfy the Jacobi identities (\ref{BianchiC}) and (\ref{BianchiB}). The superpotential (\ref{Wexample}) looks quite involved in terms of determining whether it can have  dS/Mkw vacua. 
\\[0mm]
By applying the GL(2,$\mathbb{R}$) transformation $\,\cZ=\Gamma \, U\,$ with
\beq
\Gamma \equiv \frac{1}{16^{\frac{1}{3}}} \, \left( 
\begin{array}{cr}
 1  & -1  \\
 1  &  1
\end{array}
\right)  \hspace{1cm},\hspace{1cm} \textrm{so} \hspace{1cm}  |\Gamma|^{3/2}= \frac{1}{4\,\sqrt{2}} \ ,
\label{GammaExample}
\eeq 
the algebra underlying such a flux background results in that of  $16$ in table~\ref{tablelistalgebras}. The commutators are given by those of (2.12) with $\,\kappa_{1}=1\,$, $\,\kappa_{12}=\kappa_{2}=0$,  $\,\eps_{1}=0\,$, $\,\eps_{2}=1\,$.
\\[0mm]
Applying a shift on the moduli that enter  the superpotential linearly, i.e. $\,\xi_{s}=\,-\frac{1}{2}$, $\,\xi_{t}=\,\frac{1}{2}$,
\beq
\cS= S -\frac{1}{2} \hspace{5mm} , \hspace{5mm} \cT = T + \frac{1}{2} \ ,
\eeq
we end up with the superpotential (\ref{kwModular})
\beq
\cW =  \frac{1}{4\,\sqrt{2}} \, \left[ \,  3 \, \cT \, \cZ +   \cS  - \frac{3}{2} \, \cZ^{2} - \frac{1}{2} \, \cZ^{3}  \, \right]  \ ,
\label{newW}
\eeq
where $\,\xi_{3}=\xi_{7}=\frac{1}{2}$. The tadpole cancellation conditions read $\,N_{3}=N_{7}=16$. In this very much simplified version, compare eq.~(\ref{Wexample}) to eq.~(\ref{newW}), it is easy to see, as was shown above, that the no-go theorem (\ref{lessno-go}) applies,
\beq
\Delta V \, = \, -\frac{3}{512 \, y\, \sigma \, \mu} \, < \, 0  \ ,
\eeq
and this model does not possess dS/Mkw vacua.\\

Finally we present an example of non-supersymmetric dS/Mkw vacua. Let us start with the superpotential $\cW$ in (\ref{kwModular}) and impose $\,\fg_{gauge}=\mathfrak{so(3,1)}\,$ by fixing $\kappa_{1}=\kappa_{12}=-\kappa_{2}=1$. To make the model simpler, we will also take $\,\Gamma=\mathbb{I}_{2 \times 2}\,$ as well as $\,\xi_{s}=\xi_{t}=0\,$. Hence,
\beq
\cZ=U  \hspace{1cm} , \hspace{1cm} \cS=S  \hspace{1cm} , \hspace{1cm} \cT=T \ ,
\eeq
and the superpotential reads
\beq
\begin{array}{ccc}
W(U,S,T) &=& - \, 3 \left(U^3 + U \right)\, T + \Big( \, 3 \, \epsilon_1\, U - \epsilon_1 \, U^3  + \epsilon_2 - 3 \, \epsilon_2 \, U^2 \, \Big) \, S \,\, - \\
&-& \xi_{3} \, \Big( \, \epsilon_1 - 3 \, \epsilon_1 \, U^2  +   \epsilon_2 \, U^3 - 3 \, \epsilon_2 \, U  \, \Big) + 3 \, \xi_7  \, \left( U^2 + 1 \right) \ .
\end{array}
\label{newW2}
\eeq
The original fluxes are given by $\,c_0=c_2=\tilde{c}_{2}=0 \,,\,c_1=\tilde{c}_{1}=-c_3=-1\,$ for the non-geometric $Q$ flux ;  $\,-b_{0}=b_{2}=\eps_2 \,,\, -b_{3}=b_{1} = \epsilon_1\,$ for the NS-NS flux ; $\,a_{0}=-\eps_1 \,\xi_3  + 3 \,\xi_7 \,,\, -a_{1}=a_{3}=\eps_2 \,\xi_3  \,,\, a_{2}=\eps_1 \,\xi_3  + \xi_7 \,$ for the R-R flux, and satisfy the Jacobi identities (\ref{BianchiC}) and (\ref{BianchiB}).

\begin{figure}[h!]
\centering
\includegraphics[width=14cm]{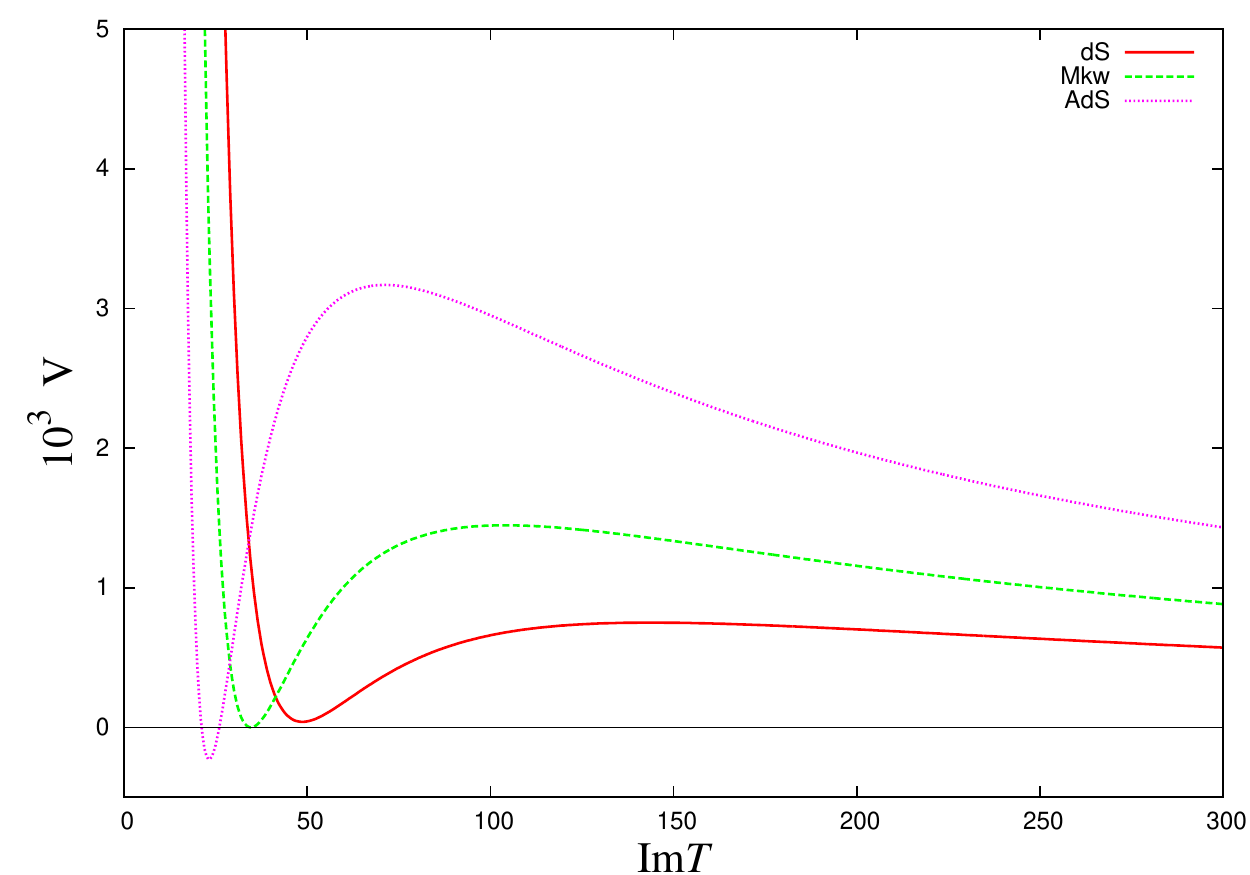}
\caption{Plot of the potential energy, $\,V\,$, as a function of the modulus Im$T$. To obtain it, we have fixed all the moduli to their VEV but the lightest one, which mostly coincides with $\,\textrm{Im}T$. The magenta/dotted line (AdS) corresponds to $\,\epsilon_{2}=45\,$, the green/dashed one (Mkw) to $\,\epsilon_{2}=44.309\,$ and the red/solid line (dS) to $\,\epsilon_{2}=44\,$. Note that a tuning of the $\,\epsilon_{2}\,$ parameter is required to obtain a Minkowski vacuum ($V_{0}=0$).}
\label{fig:V3}
\end{figure}

Further taking $\,\eps_{1}=\xi_3=1\,$ and $\,\xi_7=16$, the model is totally defined in terms of a unique $\eps_{2}$ parameter and the underlying Supergravity algebra is that of $1$ in table~\ref{tablelistalgebras}.
\begin{table}[htb]
\small{
\renewcommand{\arraystretch}{1.15}
\begin{center}
\begin{tabular}{|c|c|c|c|c|c|}
\hline
\textsc{minimum}  & $\eps_{2}$  & $U_{0}$ & $S_{0}$ & $T_{0}$ & $V_{0}$ \\
\hline
\hline
dS & $44$ & $0.435 + 0.481 \, i$ & $-1.152 + 2.008 \, i$ & $54.628 + 48.684 \, i$ & $3.983 \times 10^{-5}$ \\
\hline
Mkw & $44.3086352$ & $0.444 + 0.467 \, i$ & $-1.159 + 1.567 \, i$ & $55.084 + 34.647 \, i$ & $0$ \\
\hline
AdS & $45$ & $0.454 + 0.444 \, i$ & $-1.160 + 1.184 \, i$ & $55.897 + 23.237 \, i$ & $-2.295 \times 10^{-4}$ \\
\hline
\end{tabular}
\end{center}
\caption{Extrema of the scalar potential with positive mass for \textit{all} the moduli fields.}
\label{tableMinima}
}
\end{table}
Using the minimization procedure that will be presented in \cite{deCarlos:2009qm}, a non-supersymmetric minimum can be easily found. Moreover, as long as the $\,\epsilon_{2}\,$ parameter varies, this minimum changes from AdS to dS crossing a Minkowski point, as it is shown in table~\ref{tableMinima} and also plotted in figure~\ref{fig:V3}. 

The Supergravity algebras involving non-geometric flux backgrounds in any duality frame constitute the main set of algebra-based effective models where to perform a detailed search of dS/Mkw vacua, see table~\ref{tableexcluded}. Up to our knowledge, such an exhaustive search has not been carried out in the literature. One of the main points we would like to stress is that a plain minimisation of the scalar potential, which involves solving very high degree polynomials, is a very inefficient (and, probably, impossible) way of searching for vacua. On the other hand, an analytic calculation can be performed, using the decomposition of the scalar potential (\ref{V_Gen_NS})-(\ref{V_Gen_F}), to work out the stabilisation of the $\,\cS\,$ and $\,\cT\,$ moduli, that enter linearly the superpotential of eq.~(\ref{kwModular}). After integrating out these fields, the resulting effective potential for the $\,\cZ\,$ modulus can be tackled numerically \cite{deCarlos:2009qm}.


\section{Conclusions}
\label{sec:conclusions}

We have studied generalised fluxes in the context of the T-duality invariant ${\cal N}=1$ orientifold limits of type IIA/IIB string theory compactified on  $\,\mathbb{T}^{6}/ (\mathbb{Z}_{2} \times \mathbb{Z}_{2})\,$. Taking as a starting point the classification of all allowed non-geometric $Q$ fluxes performed in a previous paper~\cite{Font:2008vd}, we completed this task by adding the $\bar{H}_3$ flux, and considering both types of fluxes as the structure constants of the Lie algebra spanned by isometry and gauge generators, coming from the reduction of the metric and the $B$ field from 10 to 4 dimensions. We have achieved a complete classification of the 12d algebras that are compatible with the type II ${\cal N}=1$ Supergravities that the isotropic $\mathbb{Z}_{2} \times \mathbb{Z}_{2}\,$ orbifold admits. Note that, by now, we are dealing both with geometric and non-geometric fluxes. The main result of this first part is then summarised in table~\ref{tablelistalgebras}.

This task was performed in the duality frame defined by type IIB string theory with O3/O7-planes, which is where this algebra classification is at all possible, due to the absence of $\omega$ and $R$ fluxes. We then turned to the dual type IIA frame to study the recent no-go theorems that have been proposed in the literature~\cite{Hertzberg:2007wc} about the existence of Minkowski and de Sitter (Mkw/dS) vacua in ${\cal N}=1$ Supergravity models. We formulate a weaker version, see eq.~(\ref{lessno-go}), of the theorem which does not involve R-R flux, in order to combine it with the classification of algebras, which neither features this flux. 

For that purpose the next step was to perform a complete mapping of the type IIB potentials with the type IIA expressions already presented in the literature (which are given as a sum of  inverse powers of two physical moduli), including localised sources. Our expressions are complete, in the sense of including the dependence in all six real moduli, and can be rewritten in the language of ${\cal N}=1$ Supergravity, i.e. in terms of a K\"ahler potential and a superpotential.

We have finally applied the no-go theorems to these expressions (modulo some changes of basis), in order to obtain phenomenological results, i.e. whether or not the corresponding potentials would have Mkw/dS vacua. Our main result of the second part is given in table~\ref{tableexcluded}, where we establish that almost half of the possible algebras cannot accommodate Mkw/dS vacua.  Moreover, the most promising scenarios in terms of finding such vacua are those involving non-geometric fluxes. Once we have reached this stage, it is a matter of performing a dedicated search for minima of the potentials that survive the no-go theorem. This is the subject of a forthcoming publication~\cite{deCarlos:2009qm}.

To conclude, we would like to stress that the search for phenomenologically viable vacua in the context of string theory will not succeed if it is understood as the brute force task of searching for minima in a high order degree polynomial potential. We have shown that combining apparently disconnected pieces of research, such as the classification of the allowed Supergravity algebras in type IIB with the existence of no-go theorems of the presence of Mkw/dS vacua in type IIA, gives us the key to perform a systematic search for the most promising potentials. This already discards almost half of the possible scenarios, apart from giving us simplified expressions for the remaining, potentially viable ones. From now on, it will be, again, a question of looking for the right procedure in order to actually minimise the moduli potentials, which will be our next objective.

\vspace*{1cm}
\noindent
{\bf \large Acknowledgments}
\vspace*{3mm}

We are grateful to  P.~C\'amara,  A.~Font, P.~Meessen, R.~Vidal, G.~Villadoro and G.~Weatherill for useful comments and discussions. A.G. acknowledges the financial support of a FPI (MEC) grant reference BES-2005-8412. This work has been partially supported by CICYT, Spain, under contract FPA 2007-60252, the European Union through the Marie Curie Research and Training Networks "Quest for Unification" (MRTN-CT-2004-503369) and UniverseNet (MRTN-CT-2006-035863) and the Comunidad de Madrid through Proyecto HEPHACOS S-0505/ESP-0346. The work of BdC is supported by STFC (UK).

\section*{Appendix: The $\mathcal{N}=1$ isotropic $\mathbb{Z}_{2} \times \mathbb{Z}_{2}$ orientifold with O3/O7-planes}
\label{appA}
\addcontentsline{toc}{section}{\hspace{13pt} Appendix: The $\mathcal{N}=1$ isotropic $\mathbb{Z}_{2} \times \mathbb{Z}_{2}$ orientifold with O3/O7-planes}
\setcounter{equation}{0}
\renewcommand{\theequation}{A.\arabic{equation}}

Let us consider a type IIB string compactification on a six-torus $\mathbb{T}^6$, whose basis of
1-forms is denoted by $\eta^a$ with $a=1,\ldots,6$. Imposing a $\mathbb{Z}_2$ orientifold involution\footnote{In the type IIA description, the orientifold action is given by 
\beq 
\sigma \ : \  ( \eta^{1}\,,\,\eta^{2}\,,\,\eta^{3}\,,\,\eta^{4}\,,\,\eta^{5}\,,\,\eta^{6} ) \ \rightarrow \ 
( \eta^{1}\,,\,-\eta^{2}\,,\,\eta^{3}\,,\,-\eta^{4}\,,\,\eta^{5}\,,\,-\eta^{6} ) \ .
\label{osigmaA}
\eeq}, $\sigma$, acting as
\beq
\sigma \ : \  ( \eta^{1}\,,\,\eta^{2}\,,\,\eta^{3}\,,\,\eta^{4}\,,\,\eta^{5}\,,\,\eta^{6} ) \ \rightarrow \ 
( -\eta^{1}\,,\,-\eta^{2}\,,\,-\eta^{3}\,,\,-\eta^{4}\,,\,-\eta^{5}\,,\,-\eta^{6} ) \ ,
\label{osigma}
\eeq
there are $64$ O3-planes located at the fixed points of $\sigma$. We further impose a $\mathbb{Z}_2 \times \mathbb{Z}_2$ orbifold symmetry with generators acting as\footnote{There is another order-two element $\theta_3 = \theta_1 \theta_2$.}
\beqa
\theta_1 & : & ( \eta^{1}\,,\,\eta^{2}\,,\,\eta^{3}\,,\,\eta^{4}\,,\,\eta^{5}\,,\,\eta^{6} ) 
\ \rightarrow \ (\eta^{1}\,,\,\eta^{2}\,,\,-\eta^{3}\,,\,-\eta^{4}\,,\,-\eta^{5}\,,\,-\eta^{6} )  \ , \\[2mm]
\theta_2 & : & (\eta^{1}\,,\,\eta^{2}\,,\,\eta^{3}\,,\,\eta^{4}\,,\,\eta^{5}\,,\,\eta^{6} ) \ \rightarrow \ (-\eta^{1}\,,\,-\eta^{2}\,,\,\eta^{3}\,,\,\eta^{4}\,,\,-\eta^{5}\,,\,-\eta^{6} ) \ ,
\label{orbifold1}
\eeqa
implying the torus factorization of
\beq
\mathbb{T}^6=\mathbb{T}^2 \times  \mathbb{T}^2 \times \mathbb{T}^2  \,\,:\,\, (\eta^{1}\,,\,\eta^{2})  \,\,\times\,\,( \eta^{3}\,,\,\eta^{4}  ) \,\,\times\,\,( \eta^{5}\,,\,\eta^{6} )  \ .
\label{factorus}
\eeq
The full symmetry group $\mathbb{Z}_2^3$ includes additional orientifold actions $\sigma \theta_I$ that have fixed 4-tori and lead to \mbox{O$7_I$-planes}, $I=1,2,3$. 

Under this $\mathbb{Z}_{2} \times \mathbb{Z}_{2}$ orbifold group, only 3-forms with one leg in each 2-torus survive. The invariant 3-forms are
\begin{eqnarray} 
\label{3formbasis}
\begin{array}{llll} 
\alpha_{0} = \eta^{135} \;&\; \alpha_{1} = \eta^{235} \;&\; \alpha_{2} = \eta^{451} \;&\; \alpha_{3} = \eta^{613} \ ,\\ 
\beta^{0} = \eta^{246} \;&\; \beta^{1} = \eta^{146} \;&\; \beta^{2} = \eta^{362} \;&\; \beta^{3} = \eta^{524}  \ ,\\  
\end{array} 
\end{eqnarray}
which are all odd under the orientifold involution $\sigma$. 

Let us impose an additional $\,\mathbb{Z}_{3}\,$ symmetry that reflects on the isotropy of the fluxes under the exchange of the three two-tori in (\ref{factorus}). Since the ordinary NS-NS $\,H_{3}\,$ and the R-R $\,F_{3}\,$ fields are odd under the orientifold action $\sigma$, a consistent isotropic background for them can be expanded in terms of (\ref{3formbasis}).  These backgrounds read
\beq
\begin{array}{ccccccc}
\bar{H}_{135} = b_{3}  & , & \bar{H}_{235}=\bar{H}_{451}=\bar{H}_{613} = b_{2}  & \,\,\,\,,\,\,\,\, & \bar{H}_{146}=\bar{H}_{362}=\bar{H}_{524} = b_{1}  & , & \bar{H}_{246} = b_{0}   \ , \\[2mm]
\bar{F}_{135} = a_{3}  & , & \bar{F}_{235}=\bar{F}_{451}=\bar{F}_{613} = a_{2}  & \,\,\,\,,\,\,\,\, & \bar{F}_{146}=\bar{F}_{362}=\bar{F}_{524} = a_{1}  & , & \bar{F}_{246} = a_{0}  \ .
\label{HFexpand}
\end{array}
\eeq

Furthermore, an isotropic background for the non-geometric $\,Q^{ab}_{c}\,$ tensor flux with one leg in each 2-torus is also allowed by the orientifold action (see table~\ref{tableNonGeometric}).
\begin{table}[htb]
\renewcommand{\arraystretch}{1.30}
\begin{center}\begin{tabular}{c|c}
Components & Flux \\
\hline
$ Q_{1}^{35}\,,\,Q_{3}^{51}\,,\,Q_{5}^{13}$ & $\tilde{c}_{1}$ \\
\hline
$ Q_{4}^{61}\,,\,Q_{6}^{23}\,,\,Q_{2}^{45} \,\, , \,\, Q_{6}^{14}\,,\,Q_{2}^{36}\,,\,Q_{4}^{52}$ & $ c_{1}$ \\
\hline
$Q_{2}^{35}\,,\,Q_{4}^{51}\,,\,Q_{6}^{13}$ &  $ c_{0}$\\
\hline
$ Q_{1}^{46}\,,\,Q_{3}^{62}\,,\,Q_{5}^{24}$ & $ c_{3}$ \\
\hline
$Q_{5}^{23}\,,\,Q_{1}^{45}\,,\,Q_{3}^{61} \,\, , \,\,  Q_{3}^{52}\,,\,Q_{5}^{14}\,,\,Q_{1}^{36}$ & $c_{2}$ \\
\hline
$Q_{2}^{46}\,,\,Q_{4}^{62}\,,\,Q_{6}^{24}$ &  $\tilde{c}_{2}$ \\
\end{tabular}\end{center}
\caption{Non-geometric $Q$ flux.}
\label{tableNonGeometric}
\end{table}

The $Q$ and $\bar{H}_{3}$ fluxes determine the Supergravity algebra in (\ref{IIBalgebra}), so they are restricted by the Jacobi identies coming from it. In terms of the flux entries, the $Q^2=0$ constraint in (\ref{QQJacobi}) translates into
\beq
\begin{array}{ccc}
c_0 \left(c_2-\tilde{c}_2\right)+ c_1\,(c_1-\tilde{c}_1) &=& 0  \ , \\
c_2\,(c_2-\tilde{c}_2)+c_3 \left(c_1-\tilde{c}_1\right)  &=& 0  \ , \\
c_0 c_3-c_1 c_2                                          &=& 0  \ ,  
\end{array}
\label{BianchiC}
\eeq
and the $\bar{H}_{3} Q = 0$ constraint in (\ref{QHJacobi}) gives rise to
\beq
\begin{array}{ccc}
b_2 c_0-b_0 c_2+b_1(c_1- \tilde{c}_1) &=& 0 \ , \\
b_3 c_0-b_1 c_2+b_2 (c_1-\tilde{c}_1) &=& 0 \ , \\
b_2 c_1-b_0 c_3-b_1(c_2-\tilde{c}_2)  &=& 0 \ , \\
b_3 c_1-b_1 c_3-b_2 (c_2-\tilde{c}_2) &=& 0 \ . 
\end{array}
\label{BianchiB}
\eeq

The entire set of fluxes determines the general effective $\,\mathcal{N}=1\,$ superpotential $\,W\,$ computed from \cite{Aldazabal:2006up}
\beq
W=\int_{Y}  \left(  \bar F_{3}-\,S\,\bar H_{3} \,+\,Q\,\mathcal{J} \right) \,\wedge\, \Omega  \ ,
\label{WInt}
\eeq
where the contraction $\,Q\,\mathcal{J}\,$ is a $3$-form and $\,Y\,$ denotes the internal space. The $\,\mathcal{J}(T)\,$ and $\,\Omega(U)\,$ $p$-forms are the complexified K\"ahler $4$-form and the holomorphic $3$-form respectively, assuming isotropic moduli fields. A detailed derivation of the superpotential (\ref{kwiso}) can be found in the section 2 of \cite{Font:2008vd}. The K\"ahler potential for the moduli fields is given by the standard form of (\ref{kwiso}).

\end{document}